\documentclass[pre, aps, twocolumn,nofootinbib]{revtex4}

\usepackage{graphicx}

\newcommand\ˆ{\`a}

\begin{document}

\title{SOLID FRICTION \\
FROM STICK-SLIP DOWN TO PINNING AND AGING}
\author{Tristan Baumberger, Christiane Caroli}
\address{Institut des Nanosciences de Paris \cite{CNRS},
140 rue de Lourmel,
75015 Paris
(France)}
\date{\today}

\begin{abstract}

We review the present state of understanding of solid friction at low
velocities and for systems with negligibly small wear effects.

We first analyze in detail the behavior of friction at interfaces between
wacroscopic hard rough solids, whose main dynamical features are well
described by the Rice-Ruina rate and state dependent constitutive
law. We show that it results from two combined effects : ({\it i}) the
threshold rheology of nanometer-thick junctions jammed under
confinement into a soft glassy structure ({\it ii}) geometric aging,
{\it i.e.} slow growth of the real arrea of contact via asperity creep
interrupted by sliding.

Closer analysis leads to identifying a second aging-rejuvenation
process, at work within the junctions themselves. We compare the
effects of structural aging at such multicontact, very highly
confined, interfaces with those met under different confinement
levels, namely boundary lubricated contacts and extended
adhesive interfaces involving soft materials (hydrogels, elastomers).
This leads us to propose a classification of frictional junctions in
terms of the relative importance of jamming and adsoprtion-induced
metastability.

\end{abstract}
\maketitle
\section*{}
An established tradition, when writing about
solid friction, is to
date its emergence as a well identified scientific question from
Leonardo da Vinci ({\it ca} 1500). This makes it an unusually longstanding
problem, since both its fundamental physical aspects and its
modelization for the purpose of studies of {\it e.g.} seismic fault dynamics
are still under lively debate nowadays (see for example
\cite{Dawson}, \cite{MRSAmontons}, \cite{Bo}, \cite{Scholz}).
A very important progress in its phenomenologic modelling, for
macroscopic solids, was accomplished with the formulation of the
so-called {\it rate and state constitutive laws}, which emerged in the 70's
from the work on rock friction of Dieterich \cite{Dieterich}, Rice
and Ruina \cite{RR}.

It was later shown that, inspite of their simplicity in terms of
number of parameters, they provide an excellent description of most of
the salient features of the low-velocity frictional dynamics  of a
wide variety of materials, ranging from granite to paper. Such an
amazing ``universality" is naturally appealing for the physicist,
since it suggests~:

-- The possibility of a unified, largely
material-independent, description on the underlying microscopic level.

-- One step further, a possible feedback in terms of limitations
and/or refinement of the mechanical constitutive laws.

It is on this particular approach, deliberately different from that of
tribology proper
\footnote { This means, in particular, that we limit ourselves to
systems and sliding regimes such that wear and frictional
self-heating play a negligible role.}
, but which parallels those presently developed
in the fields of plasticity of amorphous materials \cite{Langer1} and
rheology of
jammed systems \cite{livre Nagel}, that we concentrate here.

\section{Introduction}
\subsection{From Amontons-Coulomb to Rate-and-State}

\begin{figure}[h]
     \includegraphics[width = 5cm]{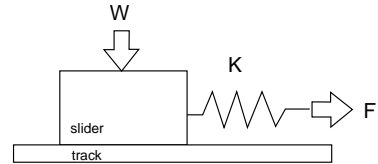}
     \caption{
     \label{fig:schema}
     A schematic slider-on-track system~: the slider is pressed onto the
     track by the normal force $W$, and pulled by the tangential force $F$
     imposed through a driving stage of stiffness $K$.}
 \end{figure}

For more than two centuries, the description of solid friction
commonly used in mechanical modelling was provided by the classical
Amontons-Coulomb laws. They state that, when a nominally planar
solid block lying on top of a planar track is submitted to a normal
force $W$ and a tangential one $F$ (Figure \ref{fig:schema})~:

- No motion occurs as long as $F$ is smaller than a finite threshold
$F_{s}$.

- Sliding is dissipative, and the corresponding dynamical friction
force $F_{d}$ is constant and equal to $F_{s}$.

- Their common value $F$ is proportional to the normal load $W$ and,
for a given $W$, independent of the macroscopic contact area
$\Sigma$. Hence, the frictional behavior of a couple of materials is
characterized by a single {\it number}, the friction coefficient~:

\begin{equation}
\label{nonumber}
\mu = \frac{F}{W}
\end{equation}

Note that this behavior (static threshold plus velocity-independent
dynamic force) is the exact analogue, for friction, of the Hill
rigid-plastic model of plasticity (threshold plus constant yield
stress), and thus probably carries unphysical singularities as is the case
of the Hill model.

Indeed, various departures from this description or its implications have
gradually emerged, the most salient of which are the following~:

(i) In general, the static friction coefficient $\mu_{s} $ is larger
than the dynamic one $\mu_{d}$.

(ii) $\mu_{s}$ is not a mere number, but a slowly {\it increasing} function
of the so-called {\it waiting time} $t_{w}$, i.e. the duration of
static contact prior to sliding.

(iii) When measured in {\it stationary sliding}, $\mu_{d}$ is not
constant. In particular, for low enough velocities (typically $<
100 \mu$m.sec$^{-1}$) it is a slowly {\it decreasing} function of $V$.

\begin{figure}[h]
     \includegraphics[width = 5cm]{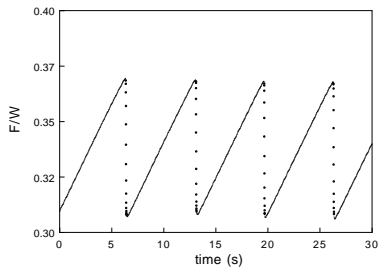}
     \caption{
     \label{fig:SS}
    Reduced tangential force {\it vs} time in the stick-slip regime for a
    paper/paper interface. From ref.\protect \cite{Heslot}.}
\end{figure}

That a stationary $\mu_{d}$ can be measured is then by itself a
puzzle, since such a velocity-weakening characteristic is well known to
make steady motion always unstable~: a positive (negative)
velocity fluctuation induces a decrease (increase) of the
friction force, leading to further accelaration (deceleration),
thereby getting amplified.
Indeed, when a slider is pulled at a low enough velocity through a driving
stage, necessarily of finite stiffness, non steady sliding is
frequently observed~: the motion is jerky, alternating between ``stick" periods
of rest during which the stage stores elastic energy, and slip events
(Figure \ref{fig:SS}). However, as skilled mechanical engineers have
known already for long, these stick-slip oscillations disappear upon
stiffening the driving stage, and steady sliding is realized.

This paradox points towards some inadequacy in the implicit assumption
underlying the above instability argument - namely that the
stationary $\mu_{d}(V)$ function can be used as such to describe non
steady motion. In other words, one suspects that the dynamic friction
force does not only depends on the {\it rate variable}, i.e. the
instantaneous sliding velocity.

\begin{figure}[h]
     \includegraphics[width = 7cm]{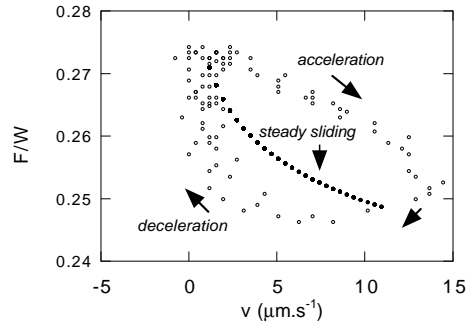}
     \caption{
     \label{fig:hysteresis}
     Hysteretic friction force response of a paper/paper system in
     non-steady sliding (open circles) and for steady sliding at various
     velocities $V$ (full circles).}
\end{figure}

(iv) Non steady friction is hysteretic, as illustrated on Figure
\ref{fig:hysteresis},
which shows the instantaneous friction force associated with a
velocity cycle for which inertia is negligible.

\begin{figure}[h]
     \includegraphics[width = 5cm]{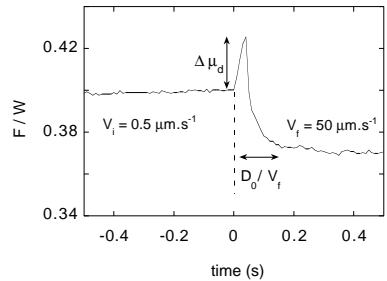}
     \caption{
    \label{fig:jump}
    Friction force transient following a jump of the driving velocity at
    $t = 0$ from $V_{i}$ to $V_{f}$, for a PMMA/PMMA interface.
    (Courtesy of L. Bureau)}
    \end{figure}

(v) Dieterich \cite{Dieterich} has studied the frictional
response to a sudden jump of the driving velocity from $V_{i}$ to
$V_{f}$ (Figure \ref{fig:jump}). He showed that it exhibits a transient the
span of
which is controlled by a characteristic {\it length} $D_{0}$, of
micrometric order. That is, its duration $\Delta t \approx D_{0}/V_{f}$.

These observations can be translated into the following statements~:

$\bullet$  A frictional interface at rest becomes stronger as time
lapses~: it strengthens when {\it aging} (see (ii)).

$\bullet$  When sliding, it becomes weaker (see (i))~: it can be said
to {\it rejuvenate upon sliding}. Moreover, its dynamic age in steady
motion decreases with increasing $V$ (see (iii)).

$\bullet$  This, together with (iv) and (v), means that its physical
{\it state} evolves with a dynamics characterized by the length $D_{0}$
and coupled to the sliding dynamics itself.

Rate and state constitutive laws model
such behaviors in terms of a single (or a few) dynamical {\it state
variable(s)}, the physical nature of which often remains unspecified.
In such a situation, the physicist's approach aims at identifying the
physical content embedded in the state variables, and at justifying on
this basis their phenomenological dynamics. One may then hope, as a
side benefit of importance, to be able to predict their
limits of validity and, possibly, to propose some further extensions.

\subsection{Spatial scales}

For this purpose, a natural first step is to identify the relevant
length scales. In the case of friction between macroscopic solids,
this immediately leads to distinguishing between two classes of
systems.\\

   (A) {\bf Rough hard solids} for which ``reasonable" loading levels
(apparent pressures $W/\Sigma$ well below elastic moduli) do not result
in an intimate molecular contact along the whole interfacial area.

For them a first relevant length scale is the average size
of the {\it microcontacts} -- the spots which form the real
area of contact $\Sigma_{r}$. It lies usually in the {\it micrometric} range.
Moreover, we will see that a detailed analysis leads to
attributing ultimately frictional dissipation to {\it elementary mechanical
instabilities} involving molecular rearrangements on the {\it
nanometric} scale.\\

   (B) {\bf Soft and/or smooth solids} which are able to get
into intimate contact everywhere along their interface. For them,
the {\it nanometric} scale retains its importance, while the
mesoscopic one becomes irrelevant.\\

In both cases, due to the long range of elastic interactions, a third
important length scale is the {\it global size of the system} $L$.
Its value has a crucial bearing on the spatial (in)homogeneity of the
frictional motion, hence on the complexity of the sliding dynamics.

An important consequence, not to be overlooked, results
straightforwardly from the identification of these spatial scales.
Reducing the ``state" of the interface to one or a few variables
necessarily implies a statistical averaging, which can be meaningful
only if performed on a large number of units. This entails that a rate
and state phenomenology of friction is legitimate only on scales
larger than a finite cutoff. In particular, when using discretized
(block) models for e.g. numerical studies of extended interface
dynamics, the basic block size must remain much larger than either the
average intercontact distance (class (A)) or at least the ultimate
nanometric scale (class (B)). That is, we contend that the question of
the regularity of the continuum limit of discretized models --- a subject
of debate in the field of mode II interfacial fracture along seismic
faults \cite{Rice-Cochard} --- though of course mathematically sound, is
physically irrelevant~:
if it should turn out that a continuum description of the fracture head
zone would imply lengths smaller than the intercontact distance, the
interface could no longer be assumed to have homogeneous frictional
properties, but should be explicily treated as a juxtaposition of
frictional (contact) and non-frictional (non-contact) regions.

\subsection{Outline}
Following Tabor \cite{Tabor}, it is both useful and physically sound
to express the friction force between two solids as
\begin{equation}
\label{eq:Tabor}
F = \sigma_{s}.\Sigma_{r}
\end{equation}
where $\Sigma_{r}$ is the real contact area, and the stress
$\sigma_{s}$ is the so-called shear strength of the interface.

Section II is devoted to friction at multicontact interfaces
between rough hard materials (MCI), an interfacial configuration
which is prevalent when dealing with macroscopic bodies.
We first analyze their geometry, then
show that the geometric factor $\Sigma_{r}$ plays, for these systems,
the part of a state-dependent variable, governed by what
we will call the {\it geometric age} $\phi$. We then show that, for MCI,
the rate dependence of the rheological factor $\sigma_{s}$ can be
assigned to local mechanical
instabilities within the nanometer-thick molecular ``junctions" forming the real
contacts. On this basis, the Rice-Ruina constitutive
law, which they originally formulated on a phenomenological basis,
results as a good approximation for the low velocity frictional behavior
of MCI. We
sketch out its consequences in terms of the sliding dynamics of a
driven spring-block system. We also point out its various limits of
validity.

These limits turn out to be of two different types.

   $\bullet$ On the one hand, within the physical framework which
identifies ``state" with geometric age, the phenomenological functional
expression of $\Sigma_{r}(\phi)$ is necessarily aproximate, and its
limits can be evaluated.

   $\bullet$ On the other hand, the detailed analysis of experiments
brings to light the limits of the above physical framework itself.
That is, it prompts the idea that a second underlying slow dynamics
manifests itself through
the rheological factor~: {\it geometric age is not the whole
story, frictional contacts also have a structural age}.

This leads us to concentrate, in
Section III, on the rheology of frictional contacts. We analyze it
for various configurations (rough-on-flat MCI, surface force
aparatus (SFA), extended soft contacts) corresponding to various confinement
and compacity levels. In all cases, it appears that {\it structural
aging/rejuvenation mechanisms are indeed at work in the contact-forming
nanometer-thick interfacial junctions}. The associated dynamics is
quite strongly system-dependent. As a guide for future investigation,
we suggest a first level of classification which distinguishes
between two main types of dissipative behaviors, namely~:

{\it (i)}
jammed junction plasticity, akin to that of soft glassy materials;

{\it(ii)} adsorption--desorption controlled dynamics.

MCI
microcontacts belong to class {\it (i)}, gel/glass contacts to
class {\it(ii)}. Both mechanisms probably contribute to friction in
SFA contacts,
while the case of elastomer/glass remains open to
discussion.

\section{Multicontact interfaces}

The renewal of interest for solid friction in the past 20 years,
triggered by the work of rock mechanicians, was aimed at modelling
friction along seismic faults. For this reason, it naturally focussed
on MCI. It has resulted in bringing to light robust features shared
by a wide variety of materials. We will see that this has led to the
building a complete framework of description, namely~:

    --- a predictive constitutive law

    --- an underlying physical interpretation opening onto further
questions concerning contact rheology (Section III).

This is why MCI friction is treated here extensively.

\subsection{Geometry of multicontact interfaces}

\subsubsection{Surface roughness}

Consider two macroscopic solids (Figure \ref{fig:schema}), referred
to as slider and
track, with nominally planar contact surfaces. Nominally means here
that they are flat on large scales comparable with the macroscopic
slider lateral size. It is well known that, due to the
unavoidable presence of steps, atomic flatness cannot be realized
over lengths much beyond the micrometric scale. An exception is
provided by lamellar solids, in particular cleaved mica for which this
range may reach up to $\sim 1$ cm. Optically smooth
surfaces, with r.m.s. roughness on the order of a few nanometers over
$1$ cm$^{2}$, are uncommon, examples being provided by float glass and
some highly polished metals. In general, natural or ground surfaces
exhibit a r.m.s. roughness between $0.1$ and a few microns on $1$ cm$^{2}$.

\begin{figure}[h]
     \includegraphics[width = 7cm]{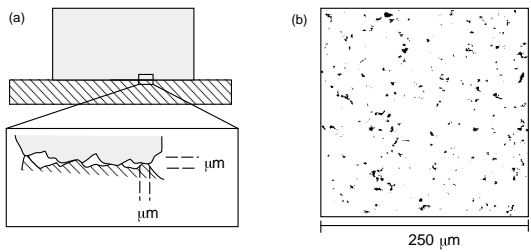}
\caption{
\label{fig:MCI}
(a) Schematic view of a multicontact interface between two nominally
planar rough solids.
(b) Optical vizualization of the light-transmitting real contacts at
the MCI between two transparent epoxy resin blocks. The contrast has been
inverted. (Courtesy of O. Ronsin)}
\end{figure}

Such random surfaces exhibit asperities with distributed heights, so
that physical contact between them occurs only at random spots, the
{\it microcontacts} between load-bearing asperities. They form what we call
multicontact interfaces (MCI) (Figure \ref{fig:MCI}).

It is Tabor \cite{Tabor}, in his pioneering work, who emphasized
the importance of distinguishing between the
apparent  and the real areas of contact, $\Sigma_{app}$ and
$\Sigma_{r}$, and of evaluating the average microcontact size
$\bar{a}$. It has emerged (see below) that typical values for
$\Sigma_{r}/\Sigma_{app}$ and $\bar{a}$ are respectively $10^{-3}$ and
a few microns. This means that a typical MCI is formed of a sparse set of
microcontacts with average separation $\simeq 100 \mu$m -- a picture
which has been confirmed by direct optical observation
(\cite{Diet-Kilg} and Figure \ref{fig:MCI}).
It is intuitively clear that increasing the normal load $W$
results in decreasing the distance between the average surface planes
(the so-called ``closure"), hence in increasing the number of
microcontacts as well as the area of preexisting ones~: $\Sigma_{r}$
increases with $W$, as illustrated by Dieterich's observations
\cite{Diet-Kilg}. Can one make the functional nature of this dependence
explicit? Does it depend on the statistical nature of the
surface roughness and if so, how? Could it be that this dependence
would explain the
Amontons proportionality between $F$ and $W$?

\subsubsection{The single microcontact}

As a first step, let us recall a few basic results from
contact mechanics concerning a single contact.
Let us focus on the a single pair of contacting asperities, modelled
as elastic spherical caps with a common radius of curvature $R$ and
Young modulus $E$, pressed together by a normal force $w$ acting along
their intercenter axis.
The problem of the resulting elastic contact was fully solved by
Hertz and is equivalent to that between a rigid plane and an elastic
sphere of radius $R^{*} = R/2$ and Young modulus $E^{*} = E/2(1-\nu^{2})$, with
$\nu$ the Poisson modulus \cite{Johnson}.

\begin{figure}[h]
     \includegraphics[width = 5cm]{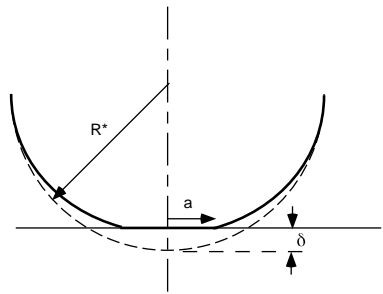}
    \caption{
    \label{fig:Hertz}
    Hertz contact between an elastic sphere and a rigid plane.}
    \end{figure}
    
Its solution can be evaluated as follows. Due to the spherical
geometry, the contact diameter $a$, the ``compression" $\delta$
(Figure \ref{fig:Hertz}) and $R^{*}$ are related by
$a^{2} \sim R^{*}\delta$. As can be checked from the full Hertz
solution \cite{Johnson},the elastic
energy is essentially stored within a depth on the order of the contact
radius, hence the relevant strain level $ \epsilon \sim \delta/a$, so that
the average normal stress $\bar p \sim w/a^{2} \sim E^{*}\epsilon \sim
\delta/a$. From this it immediately results that, dimensionally

\begin{widetext}
\begin{equation}
\label{eq:Hertz}
  a \sim \left(\frac{wR^{*}}{E*}\right)^{1/3} \,\,\,\,\,\,\,\,  \delta \sim
  \left(\frac{w^{2}}{R^{*} E^{*2}}\right)^{1/3} \,\,\,\,\,\,\,\,
  \bar p \sim \left(\frac{wE^{*2}}{R^{*2}}\right)^{1/3}
\end{equation}
\end{widetext}

The exact expressions (see \cite{Johnson}) only differ by
multiplicative constants of order unity.

Now, most solids are truly elastic only up to a yield stress $Y$, beyond
which they start deforming plastically. So, the Hertz expressions
above lose validity when the maximum normal stress below the contact
reaches this level. Upon increasing the normal load, the size of the
plastified region increases, until it occupies a volume $\sim a^{3}$.
At this stage the contact has become ``fully plastic" and deforms
so that the normal
stress remains quasi-constant~: $\bar p \simeq H$, where $H$ is the
{\it hardness}
\footnote{On the basis of geometric arguments, one estimates that $H
\simeq 3Y$\cite{Tabor hardness}.}
of the softer material. At room temperature, for metals,
the ratio $H/E$ ranges in general between $10^{-2}$
and $10^{-3}$, while for polymer glasses, $H/E \simeq 10^{-1}$ --
$10^{-2}$.
Imagine now, following Bowden and Tabor \cite{Tabor}, that the
apparent pressure $p_{app} = W/\Sigma_{app}$ is large enough for this
regime to be reached in all the microcontacts forming a MCI. Then,
the real area of contact is such that $\Sigma_{r}/\Sigma_{app} \simeq
p_{app}/H$,
and $ \Sigma_{r} \simeq W/H$.

So, in the fully plastic
regime, $\Sigma_{r}$ is proportional to the normal load, and
Amontons's law simply follows. Indeed the friction coefficient reads~:

\begin{equation}
\label{eq: fully plastic}
\mu = \frac{F}{W} = \frac{\Sigma_{r}}{W} \simeq \frac{\sigma_{s}}{H}
\end{equation}
Since, in this approximation, $\sigma_{s}$ is the shear strength under
the constant normal stress $H$, $\mu$ is effectively $W$-independent.
However rough it is, this approximation is illuminating in several
respects.

On the one hand, the a priori surprising fact that the
values of dry friction coefficients depend only weakly on the
mechanical properties of materials and are commonly a fraction of unity
can now be translated into the
statement~: interfacial shear strength values are roughly comparable
with bulk yield stress levels --- a point which will be of qualitative
importance later on.

On the other hand, it permits to get a somewhat more precise idea about
the minimum loading level necessary to form a MCI between given
materials. Consider as an example a $1$ cm thick steel plate under its
own weight. Then $p_{app} \simeq 10^{3}$ Pa, while $H \simeq 10^{9}$
Pa, hence $\Sigma_{r}/\Sigma_{app} \simeq 10^{-6}$~: a $1$ cm$^{2}$
surface would form no more than about $10$ microcontacts of area $10
\mu$m$^{2}$ (see below) -- hardly enough to form a decent statistical
set!

Tabor's suggestion, which he developed for the case of metals, later gave
rise to a number of discussions about the relevance of the full
plasticity assumption to wider classes of materials which also obey
Amontons's law while being less ductile than metals. This opened onto
the problem of modelling the elastic-plastic contact between random
surfaces.

\subsubsection{Area of contact between random surfaces}

The crucial contribution to this question was made by Greenwood and
Williamson \cite{GW}. Their model reduces the characterization of each of
the contacting
  random surfaces to~:

-- the statistical distribution of asperity summit heights
above some average plane $\phi (z)$. These asperities are assumed to
have spherical tips.

-- the asperity radius of curvature $R$, assumed to be unique.

-- the number of summits $N$ on the apparent surface $\Sigma_{app}$.

They show that the problem of contact between two such nominally flat
elastic random surfaces can be cast into that of a ``composite" random
surface and a rigid plane \cite{Johnson}.

\begin{figure}[h]
     \includegraphics[width = 7cm]{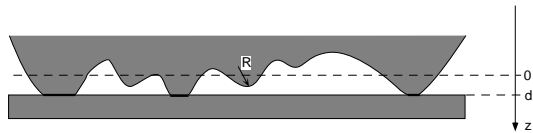}
    \caption{
    \label{fig:Greenwood}
    Schematic representation of a Greenwood-Williamson interface (see text).}
    \end{figure}
Let $d$ be the separation between the average plane and the rigid one
(Figure \ref{fig:Greenwood}), $n$ the number of microcontacts at this
separation,
assumed to be dilute enough to be treated as independent Hertz
contacts. Each asperity with height $z > d$ contributes to $n$~:

\begin{equation}
\label{eq:GW1}
n = N \int_{d}^{\infty}  dz\,\phi(z)
\end{equation}
The compression of a contact is $(z-d)$, its
area $\pi R(z-d)$, and it bears the load $(4/3)\pi R^{1/2}(z-d)^{3/2}$
(eq.(\ref{eq:Hertz}), \cite{Johnson}) so that the real area of contact

\begin{equation}
\label{eq:GW2}
\Sigma_{r} = N \int_{d}^{\infty} dz\,\pi R(z-d) \phi(z)
\end{equation}
and $d$ is related to the given total normal load $W$ by~:

\begin{equation}
\label{eq:GW3}
W = N \int_{d}^{\infty} dz\,(4/3)\pi R^{1/2}(z-d)^{3/2} \phi(z)
\end{equation}

The main physical content of the Greenwood -Williamson (GW) results
already emerges under the schematic assumption of an exponential
distribution $\phi_{0}(z) = s^{-1}\exp(-z/s)$ for $z > 0$ and zero
otherwise. One then gets~:

\begin{equation}
\label{eq:GW4}
n = \frac{1}{\sqrt \pi}\, \frac{W}{ER^{1/2}s^{3/2}}\,\,\,\,\,\,\,\,\,
\Sigma_{r} = \sqrt{\pi}\,\left(\frac{R}{s}\right)^{1/2}\,
\frac{W}{E}
\end{equation}

Both $\Sigma_{r}$ and the number of contacts are proportional to $W$,
and individual contact sizes grow under increasing load in such a way that
the average contact radius~:

\begin{equation}
\label{eq:GW5}
\bar{a} = \left(\frac{\Sigma_{r}}{\pi n}\right)^{1/2} = \sqrt{R s}
\end{equation}
remains a load-independent constant.

GW have shown numerically that these properties are conserved, to a
very good approximation, for a gaussian distribution of summit heights,
in the load range, ranging over several decades, such
that $1 \ll n \ll N$. A rough evaluation of $N$ is obtained by
representing $\Sigma_{app}$ as densely paved by average asperities of
curvature $R^{-1}$ and height $s$, so that $N \sim
\Sigma_{app}/Rs$. The upper limit then translates into

\begin{equation}
\label{eq: limsup}
\frac{p_{app}}{E} \ll
\sqrt{\frac{s}{R}}\,\,\,\,\,\,{\rm or}\,\,\,\,\,\,\Sigma_{r} \ll \Sigma_{app}
\end{equation}
Note that this condition also ensures that elastic interactions
between microcontacts can reasonably be neglected.\\

A considerable amount of work \cite{Nayak} has been devoted to
evaluating the two geometrical parameters involved in the GW model on
the basis of surface topography measurements. Curvature determinations
have been hotly debated, since they involve a second order derivative
of the profile, which makes them strongly noise-sensitive. When
measured with profilometers of lateral resolution of micrometric
order, typical values of $R$ for surfaces blasted or lapped with
abrasive powders commonly lie in the $10$ -- $100 \mu$m range
\cite{Pauline Proc Roy Soc} \cite{GW} \cite{Scholz-Engel}, while r.m.s.
roughnesses $s \sim 1 \mu$m. Then the corresponding average contact radii
$\bar{a} \sim 3$ -- $10 \mu$m, an estimate which has been confirmed
from the Dieterich-Kilgore visualizations \cite{Diet-Kilg}.

Modern experimental improvements have revealed that surface
profiles are in general more complex than was initially assumed by GW,
exhibiting multiscale roughness distributions \cite{EBouchaud}. They are often
modelled as self-affine surfaces
\footnote{The statistical properties of a self affine surface $z(x,y)$ are
invariant under the scaling transformation $x \to \zeta x, y \to
\zeta y, z \to \zeta^{3-D_{f}} z$, where $D_{f}$ is the fractal
dimension.}.

Persson \cite{Bofractal} has recently proposed a theoretical
treatment of the resulting contact problem. His calculation of the
contact area between a flat compliant medium and a rigid self-affine
surface follows a renormalization scheme for the stress distribution
for profiles at growing stages of magnification $\zeta$. This
function he proves to obey, for {\it complete} contact between the
surfaces, a Fokker-Planck-like equation, with $\zeta$ playing the part
of time. The real case of incomplete contact is then simulated  via a
boundary condition eliminating from the area of contact the regions
sustaining tensile stress. This enables him to express the solution in
terms of the elastic response of the compliant medium to imposed
displacements specified all along its surface --- while partial contact
actually corresponds to a mixed situation of given displacements within
contacts
and given (null) normal stress elsewhere. It is therefore difficult to
assess the influence of the ansatz on the main result of interest
here, namely that the real contact area still obeys the Amontons
proportionality $\Sigma_{r} \sim W$.

Note that Persson's result depends crucially on
the physical existence of a lower cut-off of the spatial
scales, which is in practice of mesoscopic order. Indeed, even in the highly
improbable case where the fresh
surfaces would remain fractal down to the atomic scale, the
nano-asperities would get plastified, and thus smoothed out, under
the very high pressures they would experience upon contact.

This remark brings up the important question of the limit of validity
of the pure elastic contact models. Clearly, when thinking in GW's
terms, plastic deformation of asperities starts coming into play when
the average pressure $\bar{p}$ becomes comparable with the yield
stress $Y$. From equations (\ref{eq:GW4}) and (\ref{eq:GW5}) , this
condition can be expressed in terms of the dimensionless
plasticity index

\begin{equation}
\label{eq:GW6}
\psi = \frac{E}{Y} \sqrt{\frac{s}{R}}
\end{equation}
which depends both on the topographic and the mechanical properties of
the interface.

The elastic regime is limited to the region $\psi < 1$, for which GW
have shown that the fraction of plastified contacts is negligible.
Persson's more elaborate theory yields a comparable evaluation, with
$s$ and $R$ in equation (\ref{eq:GW6}) now understood to be those on the
large scale (upper space cutoff). This last result takes into account
consistently the above-mentioned plastification on small lateral
scales.

On the other hand, for $\psi$ larger than a few units, most
microcontacts are in a state of full plastic deformation. Multiscale
roughness loses relevance, and one may resort to the fully plastic
version of the GW model. The matter which has flown being assumed to
redistribute on a scale much larger than contact radii (no ``piling
up"), geometry imposes that the area of a contact involving an
asperity with initial compression $(z-d)$ is $2\pi R(z-d)$. It bears the
load $\pi a^{2}H$, with $H$ the hardness. Then, trivially

\begin{equation}
\label{eq:GWplastic}
\Sigma_{r} = \frac{W}{H} \,\,\,\,\,\,\,\,\, \bar{a} = \sqrt{2Rs}
\end{equation}

 One may therefore assert the robustness of three characteristics of
multicontact interfaces, namely, to a good approximation~:

  {\it -- The real area of contact is proportional to the normal load, and
   independent of the apparent area $\Sigma_{app}$.

   -- The average contact radius is load-independent, thus introducing
   a mesoscopic length scale $\bar{a}$ defined by the topography of the
   contacting surfaces.

   -- In a wide variety of cases, $\bar{a}$ is of micrometric order.}\\

All these statements have been directly confirmed from the analysis,
by Dieterich and Kilgore \cite{Diet-Kilg}, of their optical images
of several MCI between transparent solids.

For metals, $E/Y \sim 10^{2}$---$10^{3}$. It would therefore take
unrealistically small values of $s/R$, of order $10^{-4}$---$10^{-6}$,
for metal/metal MCI to be in the pure elastic regime. They are fully
plastic, as anticipated by Bowden and Tabor.

The opposite case is that of elastomers, which are fully
(visco)elastic up to strains $\gtrsim 1$. Their interfaces, such as
that between tire and road, when analyzed in the frame of
Persson's predictions \cite{Bofractal} \cite{Bofracvisc} should provide a
test of his theory.

MCI involving polymeric glasses ($E/Y \sim 10$--$100$) pertain to the
intermediate regime, where $\psi \gtrsim 1$~: a fraction only of the
microcontacts flow significantly. Although no quantitative theory is
available, we believe that, in view of their
robustness, the above-mentioned three main results of the GW model
hold for all non-purely elastic MCI. Confirmation of this statement
has been provided by experimental investigations of various interfacial
properties, such as shear stiffness, which all exhibit the Amontons
linear dependence on normal load (see Appendix A).\\

In fine, we can summarize the above results into the following
simple statement for the friction coefficient of a MCI~:

\begin{equation}
\label{eq:GW7}
\mu = \frac{\sigma_{s}}{\bar{p}}
\end{equation}
$\bar{p}$ is the average normal stress borne by the microcontacts. It
is load-independent. Its expression in terms of the material properties
(Young modulus and yield stress) and of the topographic surface
characteristics depends upon the value of the plasticity index $\psi$
(eq.(\ref{eq:GW6})). For $\psi \gg 1$ (fully plastic regime)
$\bar{p} = H \sim 3Y$. In the opposite, fully elastic case $\psi \ll
1$, $\bar{p} \sim E\sqrt{s/R} \sim H/\psi$, and in the intermediate
elasto-plastic range it is expected to extrapolate smoothly with
$\psi$ between these two limits.

Let us insist that the interfacial shear strength $\sigma_{s}$ in
expression (\ref{eq:GW7}), which is likely to depend upon the
contact pressure, is load-independent for a given MCI, since such
is $\bar{p}$ itself. This result is specific of multicontact
interfaces, for which Amontons's law is of purely geometric origin.
It must be contrasted with the case of intimate single contacts such
as those studied with the surface force apparatus (SFA)
or extended gel glass ones (see Section III). For these systems,
Amontons's law, when observed, must be attributed to the pressure
dependence of $\sigma_{s}$ itself \cite{Robbins-500 ans}.

\subsection{Geometric age~: a major state variable for MCI }

\subsubsection{ Time dependence of the static threshold:}
The classical description of solid friction states that there exists,
for any given interface, a well defined static friction
coefficient $ \mu_{s} = F_{s}/W$ such that, as long as the applied
shear force $ F < F_{s}$, no sliding occurs.

This amounts to asserting that, at a depinning
threshold $F_{s}$, the interface commutes from a purely elastic,
reversible response to external shearing, to an irreversible,
dissipatively flowing one. Most likely, and in view of unavoidable
disorder and temperature effects, such a strong statement,
which implies that the interface would undergo an abrupt unjamming
transition, is only approximate. We come back to this point in
$\S$ II.C.4, where we discuss in detail the intrinsic difficulties
and limits associated with the definition of such  a threshold. Let
us only state at this point that, in order to minimize ambiguities and
make comparison between data meaningful, it is important to define
carefully the protocol used to measure $\mu_{s}$.

A series of such {\it stop and go} experiments consists in~:

      (i) Preparing the initial interfacial state reproducibly, by
          sliding steadily at a chosen velocity $V_{prep}$, then stopping the
          pulling.

      (ii) Waiting at rest for a given waiting time $t_{w}$ under a
          specified shear stress. This may either be self-selected by
          the system (natural arrest stress) or imposed at some value
          below this level.

      (iii) Resuming loading at a prescribed velocity $V_{load}$, for
          example $V_{load} = V_{prep}$.

\begin{figure}[h]
     \includegraphics[width = 5cm]{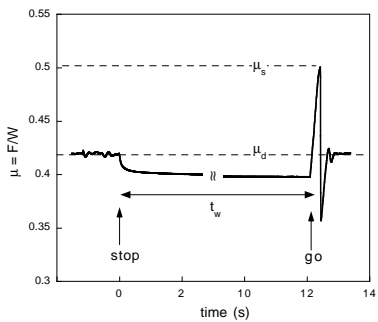}
     \caption{
    \label{fig:stop and go MCI}
    Friction trace (reduced friction force $\mu = F/W$ {\it vs} time $t$)
    in a stop and go experiment (see text). $V_{prep} = V_{load}$.
    Interface~: PMMA/PMMA at
    $T = 300$K.}
    \end{figure}

\begin{figure}[h]
     \includegraphics[width = 7cm]{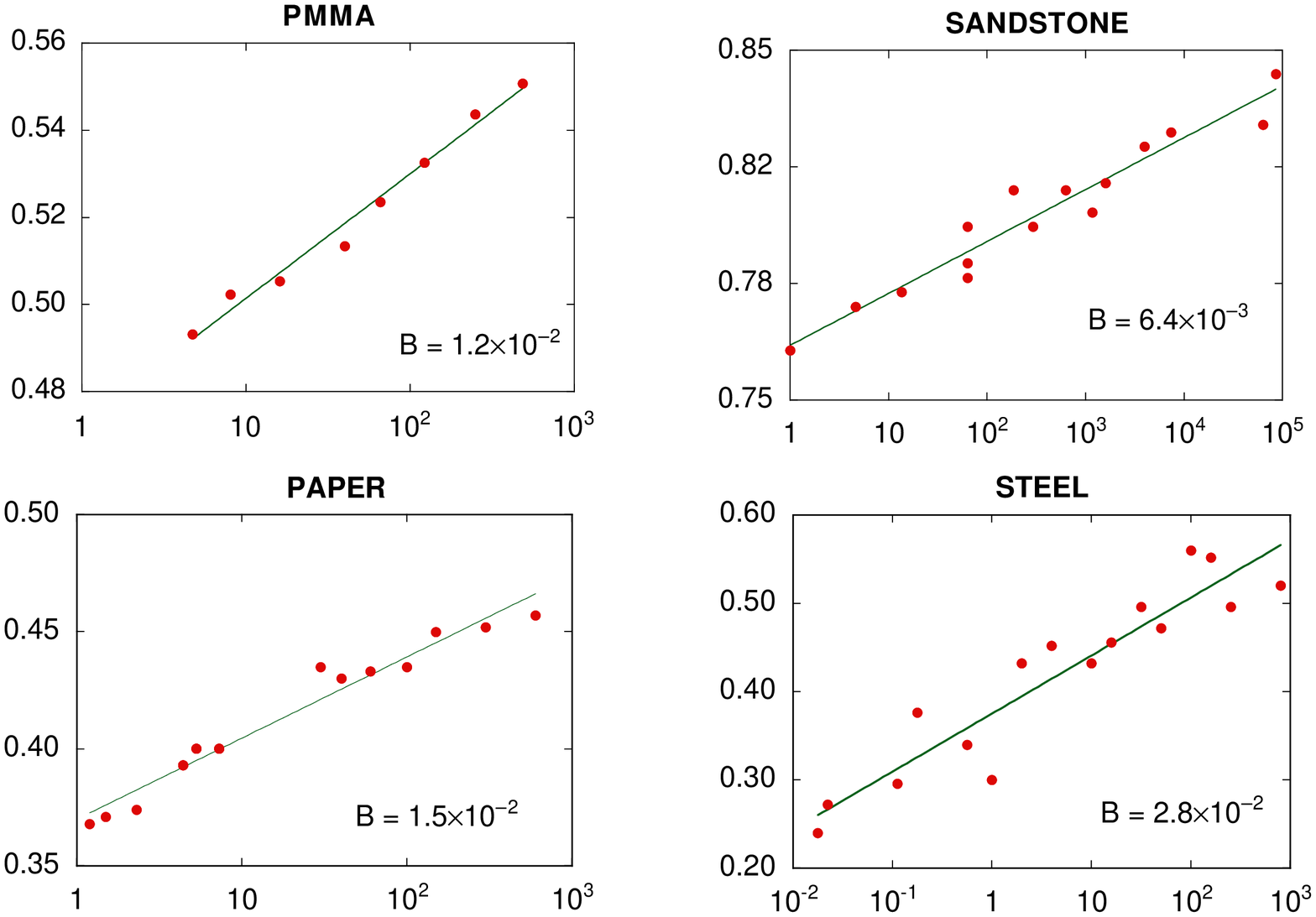}
     \caption{
    \label{fig:mus(t)}
    Static friction coefficient $\mu_{s}$ {\it vs} waiting time $t_{w}$
    for symmetric MCIs. Data from~: Heslot et al \protect\cite{Heslot} (paper),
    Baumberger \protect\cite{TBunpub} (PMMA), Dieterich \protect\cite
    {Diet-sandstone}
    (sandstone), Dokos \protect\cite{Dokos} (steel).}
    \end{figure}

A typical shear force response is displayed on Figure \ref{fig:stop
and go MCI}. $\mu_{s}$ is
then {\it conventionally} identified with the peak level.
Experiments of this type have been perfored on a number of MCI.
They reveal that $\mu_{s}$ {\it is not a mere number} characterizing a
couple of solids, but a slowly increasing function of the waiting
time. More specifically, $\mu_{s}(t_{w})$ varies logarithmically over
several decades of $t_{w}$ ranging up from about $1$ sec (the
typical fast limit for such mechanical experiments). This behavior,
displayed on Figure \ref{fig:mus(t)}, holds for a wide variety of materials
including metals \cite{Dokos}, rocks \cite{Dieterich} \cite{Marone},
glassy polymers \cite{Pauline1} and paper \cite{Heslot}.

{\it As a MCI ages at rest, it strengthens logarithmically.} Moreover,
the slope

\begin{equation}
\label{eq:age1}
B = \frac{d\mu_{s}}{d(\ln t_{w})}
\end{equation}
is found, for all the above-mentioned materials, to be roughly (see
Figure \ref{fig:mus(t)}) on the order of $10^{-2}$.

Such a ``generic" behavior is striking, and suggests that it results
from a robust physical mechanism.

\subsubsection{Creep growth  of the real area of contact}

Tabor's decomposition of the friction force (eq.(\ref{eq:Tabor})),
together with the Greenwood-Williamson analysis, immediately raises
the possibility that this strengthening might be attributable to slow
growth of the real area of contact. Indeed, we have seen that, in
general, the average normal stress $\bar{p}$ borne by the
microcontacts is comparable with the bulk yield stresses of the
contacting materials
\footnote{ More precisely, for an asymmetric interface, $\bar{p}$ is
on the order of the yield stress of the softer material.}. At this
stress level, which prevails in a volume $\sim a^{3}$ (with $a$ the
microcontact radius), one expects materials to undergo plastic creep,
resulting in the slow growth of microcontacts, hence of $\Sigma_{r}$.

Confirmation of this idea was obtained by Dieterich and Kilgore
\cite{Diet-Kilg} who were able to measure directly the time evolution
of $\Sigma_{r}$ on their optical images of the microcontacts
forming the MCI between two transparent glassy polymer blocks.
They found that $\Sigma_{r}(t_{w})$ does grow logarithmically,
at a rate compatible with that measured in microindentation experiments.

The question is then to ascertain whether or not this area growth is
sufficient to explain that of $\mu_{s}$. This has been investigated in
detail by Berthoud et al \cite {Pauline1}. They studied the
temperature dependence of the static aging slope $B$ for
symmetric MCI involving the polymeric glasses PMMA and PS
(polystyrene) between room temperature and the vicinity of the bulk glass
transitions. They analyzed their results in terms of a model for the
growth of $\Sigma_{r}$ due to Br\'echet and Estrin \cite{BE}.

This model schematizes the creep-induced growth of a microcontact as
follows. Once a microcontact has been created, at time $t = 0$,  in a
first stage of fast plastic flow, of duration negligible on the scale
of the later evolution, the normal stress $\sigma$ sets to an
"initial" value $\sigma_{Y}$ of the order of the yield stress in the
relevant geometry
\footnote{In uniaxial loading, $\sigma_{Y} = Y$, while for a
sphere-sphere contact $\sigma_{Y} = H \simeq 3Y$ \cite{Johnson}.}. Once
this state is reached, plastic evolution continues via creep, the rate
of which is given by an expression \`a la Nabarro-Herring~:

\begin{equation}
\label{eq:age2}
\dot{\epsilon} = \dot{\epsilon}_{0}\, \exp{\sigma/S}
\end{equation}
where $\dot{\epsilon}$ is the compressive strain rate (treated as a
scalar), $S$ the so-called strain-rate sensitivity of the flow stress,
and $\dot{\epsilon}_{0}$ a $\sigma$-independent Arrhenius factor.
Since this creep law reflects a thermally activated process, both
$\dot{\epsilon}_{0}$ and $S$ are $T$-dependent.

Such plastic deformation occurs at constant volume, so that~:
\begin{equation}
\label{eq:age3}
\dot{\epsilon} = \frac{1}{a_{0}^{2}} \frac{d(a^{2})}{dt}
\end{equation}
with $a$ the contact radius, $a_{0}$ its initial value. Since
$\sigma = w/\pi a^{2} \simeq (1 - \epsilon) w/\pi a_{0}^{2} +{\cal
O}(\epsilon^{2})$, with $w$
the normal load. Then, from eqs.(\ref{eq:age1}) and (\ref{eq:age2}), and
approximating the MCI by a set of identical average GW contacts, the
real area of contact evolves as~:

\begin{equation}
\label{eq:age4}
\Sigma_{r}(t) = \Sigma_{r0} \left[ 1 + m
\ln\left(1 + \frac{t}{\tau}\right) + {\cal O}(\ln^{2})\right]
\end{equation}
with

\begin{equation}
\label{eq:age5}
m = \frac{S}{\sigma_{Y}},\,\,\,\,\,\,\,\,\,\,
\tau = \dot{\epsilon}_{0}\, \frac{1}{m}\, e^{1/m}
\end{equation}

Typically, at room temperature, for the systems of interest here,
$m$ values are on the order of a few $10^{-2}$. On the other hand, no
reliable evaluation of the cross-over time $\tau$ can be made a
priori, due not only to its exponential dependence on $m$ but even
more to the lack of any precise data on $\dot{\epsilon}_{0}$.
All MCI studied up to now, except for polymer glass ones close to the
glass temperature, exhibit a linear logarithmic growth of
$\mu_{s}$, hence of $\Sigma_{r}$, for waiting times above $1$ sec.
This only enables us to state, at this stage, that in general $\tau$
is smaller than this limiting value.
Then, assuming that the Tabor interfacial shear strength $\sigma_{s}$
exhibits no aging, one obtains from eq.(\ref{eq:age4}), in the
accessible range $t \gg \tau$, for the
log-slope of the static friction coefficient~:
\begin{equation}
\label{eq:age6}
B = \mu_{s0}\,m
\end{equation}
where $\mu_{s0} = \sigma_{s} \Sigma_{r0}/W = \sigma_{s}/\sigma_{Y}$
is the static threshold
base value at short times $t \ll \tau$
\footnote{Strictly speaking the value of $\sigma_{s}$ which appears
in this expression should be that at the driving velocity (see
$\S$II.C.3 and \cite{Pauline2}}
. Since friction coefficients of
MCI are usually a fraction of unity, one thus expects, at room
temperature, $B$  to lie in the $10^{-2}$ range.

The analysis of experimental data for $B(T)$ in ref.\cite{Pauline1}
is quite intricate, in view of several difficulties concerned with~:

    -- the fact that $\tau$ and, hence, $\mu_{s0}$, could be accessed
    in the realizable waiting time range only close
    to the glass transition temperatures $T_{g}$.

    -- the problems related with mapping the authors' bulk data for
    $\sigma_{0}$ and $S$ onto the sphere-sphere geometry.

Berthoud et al were nevertheless able to show that expression
(\ref{eq:age6}) accounts semiquantitatively for the roughly tenfold
increase of $B$ between its values, of order $10^{-2}$ at
room temperature as expected from expression (\ref{eq:age6}),
and the vicinity of
the bulk $T_{g}$'s. However, it must be noted that eq.(\ref{eq:age6})
is found to systematically underestimate the experimental data, the relative
misfit increasing significantly on approaching $T_{g}$. This hints
towards the fact that, while creep-induced growth of the real area of
contact is responsible for most of the static strengthening, some
aging of the interfacial strength is not completely ruled out~: for
example, close to $T_{g}$, such symmetric polymer glass contacts might
exhibit partial ``healing" due to interdiffusion of polymer chains.\\

Clearly, the Br\'echet-Estrin model only describes {\it plastic}
creep of the contacting asperities. However, quasi-logarithmic static
aging  was also observed by Ronsin et al \cite{Coey-Rons} on rough
rubber/rough glass MCIs. For such materials, contact area growth is
obviously of viscoelastic origin. We show in Appendix B, following Hui
et al \cite{Hui}, that the Greenwood-Williamson model can be worked
out explicitly for linear viscoelastic materials, with a very
simple outcome. Namely, the GW results (eq.(\ref{eq:GW4})) for the
elastic MCI still hold formally, provided that the inverse Young
modulus $1/E$ is replaced by the so-called creep compliance $J(t)$,
which measures the delayed strain response to a unit instantaneous stress
jump. Hence the real area of contact becomes a slowly increasing
function of contact duration. A logarithmic increase is known to provide
a reasonable approximation for $J(t)$ over time decades, for materials
with a very wide spectrum of relaxation times such as rubbers.

\subsubsection{Geometric age as a dynamical state variable}

In the light of the foregoing analysis we will for the moment
attribute the strengthening of MCI static thresholds to the sole time
dependence of the area factor in Tabor's expression. $\Sigma_{r}$
thus becomes a function of the {\it geometric age of the interface} $\phi$
by which $t$ should now be replaced in equation (\ref{eq:age4}). More
precisely, this age is defined as follows.

    $\bullet$ For a non-moving MCI, $\phi(t)$ is simply the time which
has been spent at rest at time $t$.

    $\bullet$ Consider now a MCI sliding at the constant velocity $V$.
As motion proceeds, a given microcontact, once created (born) is
gradually sheared until it slides, then disappears (dies)
when the relative displacement between the partner asperities reaches
a value on the order of a fraction of the contact diameter. Since,
under constant normal load, the average number of microcontacts is
conserved, any contact death is, on average, associated with the
birth of of a new microcontact at an uncorrelated position. Since
$\Sigma_{r} \ll \Sigma_{app}$, we can safely consider that the newborn
is formed between fresh asperities, which have not yet experienced
creep
\footnote{This approximation, which neglects wear, is validated by the
observed stability of the frictional characteristics of a MCI over
slid lengths as large as a few $10$ cm \cite{unpub} -- a distance over
which we can estimate the number of contact configuration renewals
to $10^{4-5}$.}.

Contact renewal therefore limits the age of the MCI to
the average lifetime of a given configuration of microcontacts, which
can be written phenomenologically as~:
\begin{equation}
\label{eq:age7}
\phi_{ss} = \frac{D_{0}}{V}
\end{equation}
In other words, {\it  motion interrupts aging}, since
interfacial configuration
memory is destroyed after sliding the {\it characteristic
length} $D_{0}$, which we expect to lie in the micrometer range.

Note that the larger $V$, the younger the steady sliding MCI, hence the
smaller $\Sigma_{r}(\phi)$: geometric aging thus immediately appears
as a candidate process for explaining the $V$-weakening behavior of the
steady sliding dynamical friction coefficient $\mu_{d}(V)$ mentioned
in Section I.

    $\bullet$ Let us now turn to non-steady sliding at the instantaneous
velocity $\dot{x}(t)$. $\phi$ is no longer time-independent, since it
keeps track of geometric aging over the time necessary to slide the
memory length $D_{0}$. In this sense, $D_{0}$ is the length over
which memory of the history of motion is preserved.

The most simple phenomenological expression for $\phi$ accounting for
this behavior is~:
\begin{equation}
\label{eq:age8}
\phi(t) = \int_{t_{0}}^{t} dt_{1}\,\exp\left[- \frac{(x(t)
-x(t_{1})}{D_{0}}\right]
\end{equation}
where $t_{0}$ is the time at which the two solids were first brought
into contact. Note that, in agreement, as needed, with the
previously defined expressions.~:

-- for the static MCI equation (\ref{eq:age8}) yields $\phi(t) =
(t-t_{0}) \equiv t_{w}$,

-- steady sliding corresponds to the limit $t_{0} \rightarrow -\infty$,
so that $\phi(t)$ as defined from eq.(\ref{eq:age8}) reduces to the
constant $\phi_{ss}(V) = D_{0}/V$,

Geometric age thus becomes a dynamical variable, coupled to the
instantaneous velocity $\dot{x}(t)$ by the non-linear differential
equation, equivalent to expression (\ref{eq:age8}):
\begin{equation}
\label{eq:age9}
\dot{\phi} = 1 - \frac{\dot{x} \phi}{D_{0}}
\end{equation}
which was first proposed by Rice and Ruina \cite{RR}.\\

That the memory of the interfacial state of sliding MCIs is indeed
characterized by a length, a possibility first suggested by Rabinowicz
\cite{Rabino} in an often overlooked pioneer work, was established
by Dieterich \cite{Dieterich} on the
basis of his systematic exploration of frictional transients following
velocity jumps. In these experiments the slider is set into steady
motion by driving it, through a ``spring" of stiffness $K$
\footnote {More precisely, $K$ is the equivalent stiffness of the
driving stage plus slider system.}, at an
initial velocity $ V = V_{i}$. At $t = 0$,
$V$ is suddenly jumped to $V_{f}$, and one measures the force response.
As can be seen on Figure \ref{fig:jump}, this exhibits a two-step transient.

    (i) In a first, very rapid stage, for upward (resp. downward)
    velocity jumps, the instantaneous dynamic friction coefficient
    increases (resp. decreases).

    (ii) This so-called ``direct effect" is followed by a much slower
    monotonous variation in the reverse direction, ending at the
   level $\mu_{d}(V_{f})$.

Dieterich found that the duration of the transient, which is
dominated by that of stage (ii), scales as $1/V_{f}$. From this he
identified a value of $D_{0} \simeq 5 \mu$m for a granite/granite
MCI. One may then conclude that the slow part of the transient
corresponds to the gradual relaxation  of the real contact area from
its initial steady value $\Sigma_{r}(\phi = D_{0}/V_{i})$
towards its final one $\Sigma_{r}(\phi = D_{0}/V_{f})$.\\

In conclusion of this analysis, it appears reasonable to write
tentatively Tabor's expression as~:

\begin{equation}
\label{eq:age10}
F = \Sigma_{r}(\phi)\,\sigma_{s}(\dot{x})
\end{equation}
which assigns the whole state dependence to the area factor, while
assuming that the shear strength $\sigma_{s}$ only depends on the
instantaneous sliding velocity. The validity of this assumption will
be discussed at length in the forthcoming sections.

Equations (\ref{eq:age10}) and (\ref{eq:age9}), together with Newton's
equation~:

\begin{equation}
\label{eq:newton}
M\ddot{x} = F_{drive}(x,t) - F\left(\phi, \dot{x}\right)
\end{equation}
where $F_{drive}$ is the externally applied driving force, provide a
closed set of equations for the frictional motion of the slider once the
form of the functional $\dot{x}$-dependence of $F$ has been
identified.

\subsection{Junction rheology~: gross features}

\subsubsection{Junctions at multicontact interfaces}

Up to now, we have concentrated on the analysis of the contact
geometry of MCI which, though an important prerequisite, does not yet
touch upon our main question, namely that of the origin of frictional
dissipation and of the detailed nature of the associated rheology
described by the interfacial shear strength $\sigma_{s}$. Let us
first try to identify the regions which are the seat of
frictional dissipation.

\begin{figure}[h]
     \includegraphics[width = 7cm]{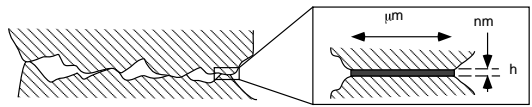}
     \caption{
    \label{fig:junction}
    Each microcontact is formed by a nm-thick adhesive junction between
    the two asperity bulks.}
    \end{figure}

We have restricted our definition of MCI to the normal load range
such that microcontacts are numerous enough to form a good statistical
set, but sparse enough for elastic interactions between them via the
bulk materials to be negligible. This entails that we can now simply
focus on the behavior of a single sheared microcontact. Such a unit
(Figure \ref{fig:junction}) is constituted of the bulk of the two
asperities and of an interfacial layer in which molecules from
both surfaces have come into adhesive contact.

This layer, which we call the {\it junction}, has a disordered structure.
This is obvious when the two solids are polymer glasses,
since in this case the junction is formed by polymer tails
and loops protruding from the amorphous bulks. In the case of
crystalline bulks, structural and chemical surface disorder, which
prevails under usual (non atomically planar and non
ultra high vacuum) conditions
 together with plastic smoothing out of nano-scale
roughness certainly lead to structural disorder.

Such highly defective structures, probably less dense than the bulks,
are likely to result in lower resistance against plastic shear
deformation. So, we will assume that {\it the junction is a disordered
quasi bidimensional medium with thickness $h$ in the nanometric
range, where shear naturally localizes.} We will show that this
assumption is borne out by the analysis of experimental data on low
velocity friction.

\subsubsection{A threshold rheology}

Stating the existence of a well-defined static friction coefficient
can be formulated equivalently as  the fact that, when the sliding
velocity $V \rightarrow 0$, $\sigma_{s}$  does not vanish, but tends
towards a finite limit $\sigma^{*}$. That is, from a rheological
viewpoint, frictional contacts behave as yield stress fluids.

{\it The yield stress $\sigma^{*}$ therefore appears as the threshold
beyond which the interfacial quasi--$2D$ solid flows plastically.}

At lower stress levels, in this simple picture, the interface does not
slide, it is pinned and responds to shear as a solid i.e., in
principle, elastically. Due to the nanometric thickness of the
junctions, the corresponding stiffness is always much larger than that
of the microcontact-forming asperities themselves (see Appendix A)
\footnote{A rough estimate for the corresponding ratio is $\bar{a}/h
\gtrsim 10^{3}$, with $\bar{a}$ the average microcontact radius, $h$
the junction thickness.}
, it is therefore not accessible in MCI configurations.

It is common knowledge that plastic flow does not set in as an
ideally sharp transition~: at non-zero temperatures, thermal
activation induces creep below the nominal yield stress, which can be
attained, in principle, only by loading at extremely high rates. In
other words, the smaller the loading rate, the more fuzzy the
threshold. So, one expects that a constant shear load close below the
static threshold should induce creep-like frictional sliding. This
was indeed observed by Heslot et al \cite{Heslot}.
\footnote{Note, however, that with MCI this creep is certainly
amplified by the geometric rejuvenation (weakening) associated with
sliding.}.

\subsubsection{Beyond threshold~: rate effects}

\paragraph{Velocity jumps~: the direct effect.}

We already mentioned in $\S$II.B.3 that the force response (see
Figure \ref{fig:jump}) to a jump of the driving velocity from $V_{i}$ to
$V_{f}$
exhibits, previous to the slow transient attributable to geometric age
adaptation, a first much faster part. The associated force jump $W
\Delta \mu_{d}$ is positive (resp. negative) for $V_{f} > V_{i}$
(resp. $V_{f} < V_{i}$), hence the term ``direct effect". Dieterich has
shown that~:

\begin{equation}
\label{eq:jump}
\Delta \mu_{d}(V_{i} \rightarrow V_{f}) = A \ln
\left(\frac{V_{f}}{V_{i}}\right)
\end{equation}
where $A$ is a constant for a given couple of materials. Its values
lie in the $10^{-2}$ range.
This holds for all the MCI studied up to now, for velocities between
about $0.1$ and $100 \mu$m.sec$^{-1}$.

Note that, as long as inertia is negligible, motion is quasi-static,
i.e. after the jump at $t = 0$~:

\begin{equation}
\label{eq:jump2}
K \left(V_{f}t - x\right) = W \left[\mu_{d}\left(\phi,
\dot{x}\right) - \mu_{d}^{st}\left(V_{i}\right)\right]
\end{equation}

$x(t)$ measures the position of some reference point on the slider
along the pulling direction, $x(0) = 0$, and $\mu_{d}^{st}(V) =
\mu_{d}\left(D_{0}/V , V\right)$ is the friction coefficient for
steady sliding at velocity $V$.

The peak value of the transient force ($d\mu_{d}/dt = 0$) therefore
occurs for $\dot{x} = V_{f}$. Velocity jump experiments are performed
with as stiff as possible driving stages such that, as can be checked
by direct measurements, the distance slept during the fast part of the
transient be much smaller than the memory length $D_{0}$. Under such
conditions the corresponding variation of geometric age from its
initial value $\phi_{i} = D_{0}/V_{i}$ is negligible, and the direct
effect is fully attributable to the rate dependence of the shear
strength $\sigma_{s}$.

From equation (\ref{eq:age10}) one then expects that~:

\begin{equation}
\label{eq:jump3}
\Delta\mu \left(V_{i} \rightarrow V_{f}\right) =
\frac{\Sigma_{r}(D_{0}/V_{i})}{W} \left[ \sigma_{s}(V_{f}) -
\sigma_{s}(V_{i})\right]
\end{equation}

That is, making use of equations (\ref{eq:age4}) and (\ref{eq:jump})

\begin{equation}
\label{eq:jump4}
\Delta \sigma_{s} = \sigma_{s}(V_{f}) - \sigma_{s}(V_{i}) = \frac{A
\ln \left(V_{f}/V_{i}\right)}{\Sigma_{r0} \left[ 1 + m \ln \left( 1 +
\frac{D_{0}}{V_{i}\tau}\right)\right]}
\end{equation}

Since $m \sim 10^{-2}$, the log term in the denominator can be
neglected, and $ \Delta \sigma_{s} \simeq \frac{AW}{\Sigma_{r0}}
\ln\left(\frac{V_{f}}{V_{i}}\right) = A \sigma_{Y}
\ln\left(\frac{V_{f}}{V_{i}}\right)$, from which we can write the
following empirical expression for $\sigma_{s}$~:

\begin{equation}
\label{eq:sigmas}
\sigma_{s}(\dot{x}) = \sigma_{s0} \left[ 1 + \alpha \ln
\frac{\dot{x}}{V} + {\cal O} (\ln^{2}) \right]
\end{equation}
with

\begin{equation}
\label{eq:alpha}
\alpha = \frac{A \sigma_{Y}}{\sigma_{s0}}
\end{equation}
and $\sigma_{s0} \equiv \sigma_{s}(V_{0})$ is the shear strength at
the reference velocity $V_{0}$, which may be chosen anywhere in the
range ($0.1$ -- $100 \mu$m/sec) where equation (\ref{eq:jump}) holds.

The necessity of resorting to an expression involving a finite
reference velocity is imposed by the formal divergence of the
logarithm in the vanishing $\dot{x}$ limit. This divergence is
of course unphysical and only points out the limits of the empirical
approach. We will return to this point later, when discussing the
behavior to be expected for $\sigma_{s}(\dot{x})$ in the very low
and large velocity regimes.

Note finally that $\sigma_{s0}/\sigma_{Y}$ is the friction
coefficient $\mu_{s0}$ appearing in equation (\ref{eq:age6}), it is
therefore roughly of order $1$ and, in order of magnitude~:

\begin{equation}
\label{eq:alpha2}
\alpha \sim A \sim 10^{-2}
\end{equation}

\paragraph{Steady sliding friction coefficient.}

The velocity jump experiments initiated by Dieterich have played a
pioneering role to evidence the state and rate character of MCI
friction and to separate clearly the age (state) and rheologic (rate)
contributions to the dynamic friction coefficient $\mu_{d}$. However,
it is the studies of the velocity and temperature dependences of
$\mu_{d}(V)$ under steady sliding conditions which have led to a
systematic confirmation of the validity of expression
(\ref{eq:sigmas}) and opened the way to its more quantitative
explicitation.

\begin{figure}[h]
     \includegraphics[width = 7cm]{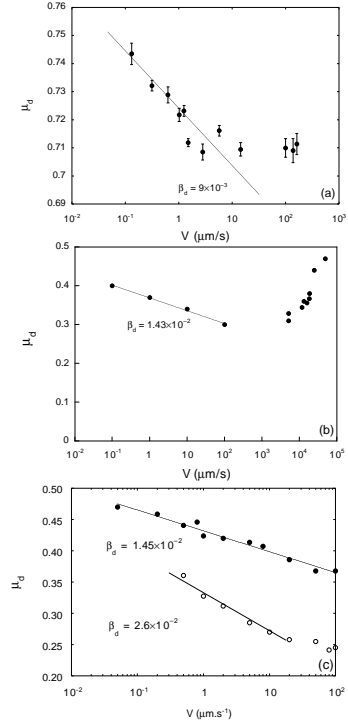}
     \caption{
    \label{fig:mud}
    Dynamic friction coefficient $\mu_{d}$ versus steady sliding velocity
    $V$ for symmetric MCI~: (a) granite (data from
    \protect\cite{Dieterich-mud}) (b) paper(data from
    \protect\cite{Heslot}) (c) PMMA at $T = 296^{\circ}$K (full symbols)
    and $T = 384^{\circ}$K $\simeq T_{g}$ (empty symbols) (data from
    \protect\cite{Pauline2}). Due to the increase of the crossover time
    $\tau$ (see eq.(\ref{eq:age5})) with $T$, close below $T_{g}$, $V_{min}$
    has decreased enough to lie in the experimental range.}
    \end{figure}

The typical experimental $V$-dependence of $\mu_{d}$ is illustrated on
Figure \ref{fig:mud}, on the cases of granite, paper and PMMA symetric MCI
at room
temperature. It is seen that~:

    (i) $\mu_{d}$ is $V$-weakening ($d\mu_{d}/dV < 0$) in the whole
low velocity regime explored in such experiments.

    (ii) Its variations are quasi-logarithmic, except in the higher
$V$ range ($V \sim 100 \mu$m/sec) where it exhibits a saturating trend.
For all materials, the log-slope  $\beta_{d} = - d\mu_{d}/d(\ln V)$
lies, once more, in the $10^{-2}$ range.

Such results can be confronted with the predictions obtained from
equations (\ref{eq:age10}), (\ref{eq:age4}) and (\ref{eq:sigmas})
which yield, in steady motion where the geometric age $\phi =
D_{0}/V$~:

\begin{eqnarray}
\label{eq:mud}
\mu_{d}(V) & = & \frac{\sigma_{s0}}{\sigma_{Y}}\,\left[ 1 + m \ln\left(1+
\frac{D_{0}}{V\tau}\right) + \alpha \ln\frac{V}{V_{0}}
\right]\nonumber\\
& = & \frac{\sigma_{s0}}{\sigma_{Y}} + B \ln\left(1+
\frac{D_{0}}{V\tau}\right) + A \ln\frac{V}{V_{0}}
\end{eqnarray}
where we have taken advantage of the smallness of $m$ and $\alpha$ to
neglect the ${\cal O} (\ln^{2})$ terms, and made use of equations
(\ref{eq:age6}) and (\ref{eq:alpha}).

For $ V \ll D_{0}/\tau$, equation (\ref{eq:mud}) gives for the
log-slope of $\mu_{d}$ the value~:

\begin{equation}
\label{eq:betad}
\beta_{d} = B - A
\end{equation}

The fact that $\beta_{d}$ is positive entails that, {\it at room
temperature}, $B > A$. Moreover, the $\beta_{d}$ data are found to be
quantitatively compatible with the available independently measured
$B$ and $A$.

On the other hand, the saturating trend at the larger $V$'s appears as
resulting from that of the real area of contact in the small age
limit, which becomes relevant for $ V \gtrsim D_{0}/\tau$. More
precisely, equation (\ref{eq:mud}) predicts a broad minimum of
$\mu_{d}$, which should occur for $V_{min} =
(D_{0}/\tau)(B - A)/A \approx D_{0}/\tau$.

That is, expression (\ref{eq:mud}) predicts that, beyond $V_{min}$,
dynamic friction should become {\it velocity-strengthening}. This
natural physical consequence of the nature of geometric aging, which
has been checked experimentally on the case of paper \cite{Heslot},
has been overlooked up to now in most studies of fault dynamics. It is
important in view of its bearing upon the nature of the sliding dynamics.

We have seen in $\S$II.B.2 that parameter $B$ varies with temperature
in a way which is
explained by its physical origin -- the thermally activated creep
of $\Sigma_{r}$. On the same polymer glass systems, Berthoud et al
have shown \cite{Pauline2} that $A$ increases quasi-linearly
\footnote{Note, however, that the T-range in these experiments is small.}
when increasing $T$ up to the vicinity of $T_{g}$. Blanpied et al
\cite{Blanpied} and Nakatani \cite{Nakatani} have investigated the
behavior of, respectively, $\beta_{d}$ and $A$ for granite over
wide $T$-ranges, from room temperature to $800^{\circ}C$ (still well
below the melting point). $\beta_{d}$ was found to decrease with
increasing $T$, enough for hot granite to become
velocity-strengthening -- a point of importance in the context of
deep seismicity. $A$, on the contrary, increases in a quasi-linear
manner. We defer a more detailed analysis and interpretation of these
results, in particular those concerned with $A$, to $\S$ II.C.4
below, where we propose a physical model for the rate effect.

\subsubsection{Threshold rheology as a signature of multistability}

The primary feature of frictional rheology is the existence of a
threshold. That is, the force $F$ needed for an interface to slide at
velocity $V$ remains finite for vanishing $V$'s. This is obviously at
odds with the standard picture of dissipation at low rates, namely~:
if the externally imposed rate of shear is much smaller than the
internal relaxation ones, the system evolves {\it
quasi-adiabatically}, so that dissipation vanishes linearly with $v$.
As first formulated by M. Brillouin in 1904 \cite{Brillouin}, in
order for $F$ to exhibit a finite threshold, it is therefore
necessary that, however slow the drive, the sheared medium evolves
through a succession of adiabatic adaptation periods interspeded with
fast instability events. Each of these events then corresponds, even
at vanishing $v$, to a finite energy loss, thus accounting for the
threshold behavior.

In modern language, {\it a threshold rheology implies multistability},
as has been developed at length in various fields such as magnetic
hysteresis, wetting dynamics, charge density wave and type-II
superconducting transport \cite{Fisher}.\\

\paragraph{A toy model for junction rheology}\

\begin{figure}[h]
     \includegraphics[width = 7cm]{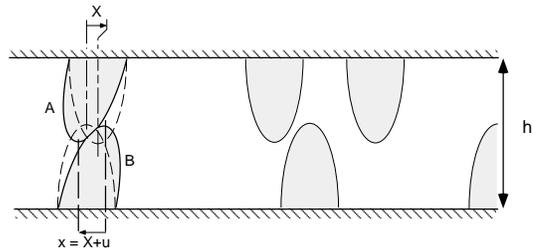}
\caption{
\label{fig:toy}
(a) Toy model of the adhesive junction. $X$ denotes the relative
displacement of blob {\bf A} with respect to {\bf B} in the absence
of shear compliance ($k = 0$), $x = X + u$ the real relative coordinate
of the shear-deformed {\bf A}/{\bf B} contact.  (b) The repulsive
interblob potential $U(x)$.}
\end{figure}

Let us first illustrate this idea here on a toy model of a sheared junction
\footnote{This model was first developed by Caroli and
Nozi\`eres \cite{CNTrieste}. Although their formal results directly
apply here, let us point out that the original physical
interpretation -- that the elementary mechanically
unstable units were the interasperity microcontacts as a whole --
was not correct \cite{CNinter}.
Let us insist again that {\it the relevant instabilities  do not occur on the
micrometric scale but, within the inter-asperity junction, on the nanometric
scale}.}.
We represent it, as shown on Figure \ref{fig:toy},
as formed by a set of identical blobs of elastic matter attached to two stiff
plates with vertical separation $h$, and area $\Sigma$. The blobs are randomly
distributed along the surfaces of these plates, and form $n$ contacts
per unit area. They compress each other when making contact, in which
case they interact via the (repulsive) pinning potential $U(x)$, where $x$
measures their relative position along the
shearing direction  (Figure \ref{fig:toy}). As $x$ is increased,
some contacts are being destroyed while others are created at random
positions, so that $n$ remains constant. We assume, for simplicity,
that horizontal displacements are purely one-dimensional.

The system is sheared by imposing to the upper plate the
displacement $X$ with respect to the lower one. Let $k$ be the shear
stiffness of the system formed by two contacting blobs. Under the
action of the pinning potential,  the center of each blob contacting
surface undergoes an elastic shear displacement $u/2$ along $Ox$, so
that the horizontal separation $ x = X + u$. The total energy of a contact is
thus~:

\begin{equation}
\label{eq:energy}
{\cal E}(X, u) = U(X+u) + \frac{1}{2}\, ku^{2}
\end{equation}

Assume for the moment that $T = 0$. Then, for fixed $X$, $u$ sets at the value
$u^{*}(X)$ which minimizes ${\cal E}$, such that~:

\begin{equation}
\label{eq:equil}
U'(X + u^{*}) = -\, ku^{*}
\end{equation}

Two cases must then be distinguished, depending on whether the
instantaneous equilibrium thus defined is unique (monostable contact)
or not (bistable contact).\\

\begin{figure}[h]
     \includegraphics[width = 5cm]{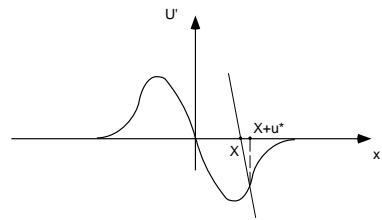}
     \caption{
    \label{fig:equilmono}
    Graphical solution of the local equilibrium equation (\ref{eq:equil})
    in the monostable case.}
    \end{figure}

{\bf Monostable contact}:

If $k > k_{0} = max( - U'') \sim U_{0}/a^{2}$,
-- with $a$ the range of $U$, $U_{0}$ its maximum value -- the solution of
eq.(\ref{eq:equil}) is unique whatever the value of the reference
coordinate $X$ (see Figure \ref{fig:equilmono}).

Assume that we impose a constant external shear velocity $\dot{X} =
V$. As $X$ increases, $u$ nearly follows its instantaneous equilibrium
value $u^{*}(X)$, and the contact energy evolves as ${\cal E}^{*}(X)
\equiv {\cal E} \left(X, u^{*}(X)\right)$. Indeed, the excess elastic
energy is dissipated out of the contact region of size $\sim a$ via
acoustic radiation, at the characteristic rate $\tau_{eq}^{-1} \sim
c/a$, with $c$ a sound velocity, hence much
larger than the shear rate $\dot{\gamma} \sim V/h \sim V/a$.

The instantaneous pinning force exerted by the contact on the upper
plate is simply~:

\begin{equation}
\label{eq:pin1}
f_{p} = - \frac{d{\cal E}^{*}}{dX} = - \left.\frac {\partial{\cal E}}{
\partial{X}}\right\vert_{u = u^{*}} - \left.\frac {\partial{\cal E}}{
\partial{u}}\right\vert_{u = u^{*}}\, \frac{du^{*}}{dX}
\end{equation}
i.e., taking advantage of eq.(\ref{eq:equil})

\begin{equation}
\label{eq:pin2}
f_{p}(X) = - U'\left(X + u^{*}(X)\right)
\end{equation}
and the net work spent to sweep through the contact~:

\begin{equation}
\label{eq:travailmono}
w = - \int_{- \infty}^{\infty} dX\,f_{p}(X) = \left.U\left(X +
u^{*})\right)\right \vert_{-\infty}^{\infty} = 0
\end{equation}
The total instantaneous friction force $F(V)$ for a junction of
area $\Sigma$ is the sum of the pinning forces $f_{p}$ over the
$n\Sigma$ randomly positioned contacts of transverse range $a$, so
that~:

\begin{equation}
\label{eq: Fmono}
F = \frac{n\Sigma}{a} \int_{-\infty}^{\infty}dX\,\left(-
f_{p}(X)\right) = n\Sigma\frac{w}{a} = 0
\end{equation}
As could be expected, {\it for an elastically monostable junction,
frictional dissipation vanishes when $V \rightarrow 0$}.\\

{\bf Bistable contact}~:

\begin{figure}[h]
     \includegraphics[width = 7cm]{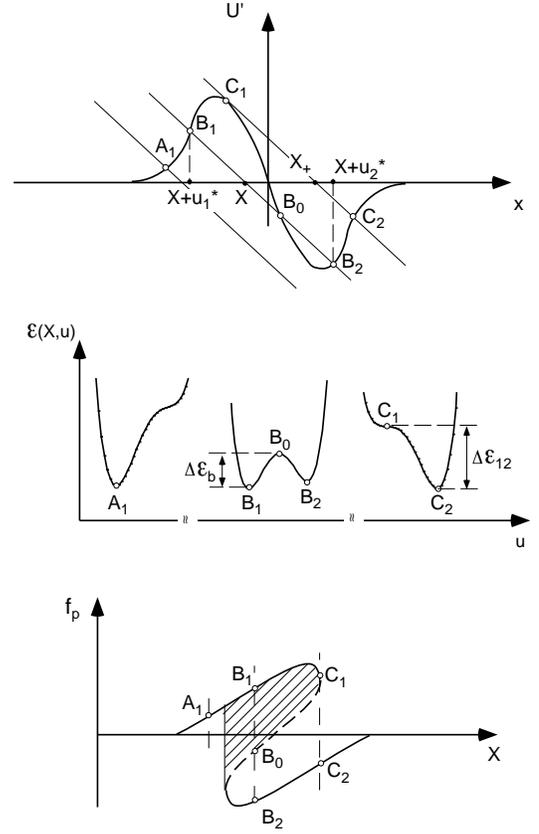}
    \caption{
    \label{fig:equilbi}
    (a) Local equilibrium construction for a bistable contact for various
    values of the reference coordinate $X$.  (b) Evolution of the contact
    energy ${\cal E}(X,u)$ for increasing $X$. The extrema conrrespond to
    the intersections labelled in (a). (c) Pinning force cycle for the
    bistable contact. When driven towards increasing $X$, the system
    follows branch ($1$) up to $C_{1}$, where it jumps down to $C_{2}$.}
    \end{figure}
    
In the opposite case where $k <
k_{0}$, in a finite range of values of $X$ equation (\ref{eq:equil})
has three solutions $u_{i}^{*}(X)$ (Figure \ref{fig:equilbi}), the two
extreme ones $u_{1,2}^{*}$ corresponding to minima ${\cal
E}_{1,2}^{*}(X)$ of the energy ${\cal E}(X,u)$, while $u_{0}^{*}$ is
associated with a maximum. As $X$ is slowly increased from $-\infty$,
the instantaneous $u^{*}$ evolves continuously with ${\cal E}$, i.e.
follows branch
$u_{1}^{*}$ (Figure \ref{fig:equilbi}b). The ${\cal E}_{1}^{*}$
equilibrium gradually evolves from fully to meta-stable, while the barrier
$\Delta{\cal E}_{b}(X)$ separating the two basins of ${\cal E}$ decreases
until the
spinodal limit $X = X_{+}$ where it vanishes. At this bifurcation
point, branch $(1)$ terminates, $u_{1}^{*}$ becomes unstable, and the
only choice for the system is to relax towards the, now single, lower
minimum ${\cal E}_{2}^{*}$. The corresponding {\it finite} energy
difference $\Delta{\cal E}_{12}$ is, again, lost via acoustic radiation, i.e.
instantaneously on the scale of the drive. The pinning force $f_{p}$
exhibits the hysteresis cycle common to such bistable systems
(Figure \ref{fig:equilbi}c). The work for sweeping through the contact
(the hatched area noted ($1$) on Figure \ref{fig:equilbi}c) $w =
\Delta{\cal E}_{12}$ is now finite.
    
\begin{figure}[h]
     \includegraphics[width = 7cm]{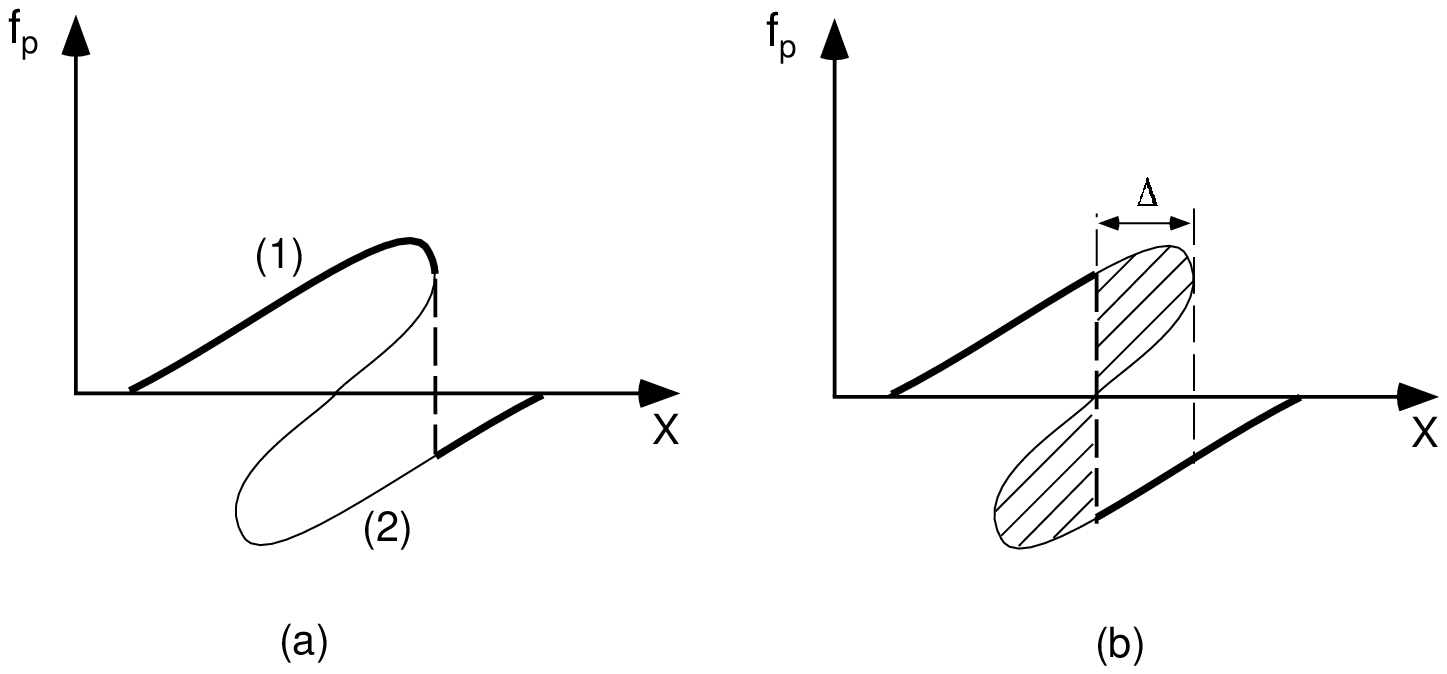}
     \caption{
    \label{fig:recoil}
    The thick lines correspond to the populated parts of the force cycle
    for (a) the sliding junction (b) the junction at rest after turning
    off the shearing force. $\Delta$ is the associated rigid recoil.}
    \end{figure}

The friction force F on the whole junction corresponds to a random
distribution of $X$, i.e. to a uniform population $P(X) dX = n\Sigma
dX/a$ of branch ($1$) of the hysteresis cycle (Figure \ref{fig:recoil}a),
so that
$F_{d} = n\Sigma w/a$. {\it Multistability results in a threshold
rheology} and, as long as $V \ll c$, the
dynamic friction force is velocity independent
\footnote{The corrections resulting from imperfect adiabatic
adaptation are easily shown to be of relative order $(V/c)^{2/3}$
\cite{CNTrieste}}.\\

Let us now assume that, starting fom this steady sliding regime, we
suddenly suppress the external shear force. The interblob contacts must
react so as to bring the system to global mechanical equilibrium,
where the sum of the pinning forces vanishes. The only way for the
junction to realize this situation is via a recoil $\Delta $
of the upper plate~: each individual reference coordinate $X_{i}$ decreases to
$(X_{i} - \Delta )$, yielding a new distribution $P_{eq}(X)$ such
that the two shaded areas on Figure \ref{fig:recoil}b be equal.
$P_{eq}$ has a discontinuity at the Maxwell plateau of the force
cycle. If, starting from this state, we now gradually increase the
external shear force $F$, $P_{eq}$ rigidly shifts to the right by
$\delta X(F)$. As long as $\delta X < \Delta$ , no irreversible ``jump"
occurs, the
displacement is therefore fully reversible~: the junction is in an elastic
regime. This terminates when the discontinuity of the shifted distribution
reaches the spinodal limit $X_{+}$, i.e. when $\delta X = \Delta$, i.e.  $F
= F_{d}$.
In this stop and go scenario, the static threshold force is equal to the
dynamic one.

However, due to multistability, global equilibrium can be realized by
a huge number of configurations of the interblob contacts.
Let us
assume that we create the unsheared junction by bringing the two plates into
contact. Depending on the details of this ``mechanical quench", each
newly formed interblob contact may correspond to a state on either of
the two branches of the force cycle, the only condition to be
satisfied being that~:

\begin{equation}
\label{eq:arret}
\int dX \sum_{i = 1,2} P_{i}(X) f_{pi}(X) = 0
\end{equation}
where $i$ labels the branches of the force cycle. The elastic regime
ends when the edge of the shifted $P_{1}$ reaches $X_{+}$. This
point, where irreversible sliding starts, is the static threshold. Contrary to
dynamic friction, {\it it is not an intrinsic property of the junction but
depends on its past history}. If, for example, $P_{1}(X_{+}) \neq 0$,
dissipation starts at vanishingly small shear.
This has been illustrated in ref.
\cite{jamming} where it was shown that the dispersion in measured
values of $\mu_{s}$ for a PMMA MCI is strongly reduced by preparing
the initial state in a controlled way, i.e. by sliding before
repose.\\

{\bf Rate effect at finite temperature~:}

Consider a bistable contact,
swept along branch $1$ of the force cycle, i.e. sitting at the left
minimum of the ${\cal E}(u)$ potential (Figure \ref{fig:equilbi}.
As soon as $T \neq 0$, thermal noise is able to activate jumps over
the barrier $\Delta{\cal E}_{b}(X)$, i.e. to provoke {\it premature
jumps} onto branch $2$ before the spinodal limit $X_{+}$ is
reached. For fixed $X$, the jumping rate is the Kramers one~:

\begin{equation}
\label{eq:rate}
\frac{1}{\tau(X)} \approx \omega_{a} \exp \left[ - \frac{\Delta{\cal
E}_{b}(X)}{k_{B}T}\right]
\end{equation}
with $\omega_{a}$ an attempt frequency, here typically $\approx c/a$.
As $X$ approaches $X_{+}$, $\Delta{\cal
E}_{b}$ decreases smoothly to zero, and $\tau(X)$ decreases exponentially up
to a cut-off fixed by viscous (acoustic) dissipation. In
practice, only close below the spinodal limit does the activated jumping
rate become non negligible. The steady sliding distribution $P$ is
now controlled by the competition between~:

    -- the advection imposed by the external drive of $X$ at velocity
    $V$,

    -- the thermally induced jumps from branch $1$ to branch $2$.

This is expressed by the evolution equation
\footnote{This expression assumes implicitly that the drive is slow
enough for the instantaneous jumping rate to be that for the non-advected
system}:

\begin{equation}
\label{eq:evol}
\frac{\partial P}{\partial t} = - V\,\frac{\partial P}{\partial X} -
\frac{P}{\tau (X)}
\end{equation}

the steady solution $P_{st}$ of which reads~:

\begin{equation}
\label{eq:distrib}
P_{st}(X) = Const\,\exp\left[- \int_{-\infty}^{X} \frac{dX'}{V \tau(X')}\right]
\end{equation}

Due to the exponential variation of $\tau$, $P_{st}$  is quasi-
completely depleted between $X_{+}$ and a cut-off  $X_{c}$ below which
activation plays a negligible role. The friction force
is thus smaller than its $T = 0$
limit. Moreover, the larger $V$, the less time the contact spends in the
vicinity of a given $X$, the less the probability for it to jump
prematurely. Hence, $X_{c}$ increases with the driving velocity, and
the friction force increases accordingly~: {\it thermal noise results in a
velocity-strenghtening rheology}.

This rate dependence can be evaluated analytically as long as the
the barrier height evolution for $X \sim X_{c}$ can be approximated simply.
For not too large $V$, where variations are linear, it is found
(see ref. \cite{Bo}, Chapter 11) that~:

\begin{equation}
\label{eq:fric-activ}
F(V) = F(V_{0})\,\left[1 + \alpha_{th} \ln\left(\frac{V}{V_{0}}\right)\right]
\end{equation}
with $V_{0}$ is a reference velocity in the above-mentioned range of
validity,and~:

\begin{equation}
\label{eq:alphath}
\alpha_{th} = \frac{k_{B}T}{\bar{\sigma} v_{act}}
\end{equation}
and $\bar{\sigma} \simeq nw/a$ is on the order of the $T = 0$
frictional stress.
  At larger velocities, when $X_{c}$ comes very close below
$X_{c}$, $\Delta{\cal E}_{b} \sim (X_{c} - X)^{3/2}$, and the rate
dependence of $F$ commutes from log-linear to $(\ln V)^{2/3}$
\cite{CNTrieste} - a behavior which has been observed by Sills and Overney
\cite{Overney} in a nanoscale friction experiment on glassy
polystyrene performed with an atomic force microscope.\\

\paragraph{From the toy model to the real junction~:}\

When noticing that the functional form predicted by the
toy model (eq.(\ref{eq:fric-activ})) for the  interfacial rheology
does fit the empirical expression (eq.(\ref{eq:sigmas})) deduced from
experiments, one is led to go one step further. That is, let us take
for a moment the toy model at face value and compare experimental
results for coefficient $\alpha$ (eq.(\ref{eq:alpha})) with
$\alpha_{th}$ (eq.(\ref{eq:alphath})). As mentioned above, Nakatani
\cite{Nakatani}
found that, for granite, $\alpha$ increases quasi-linearly with
temperature, in agreement with equation (\ref{eq:alphath}) -- a result
confirmed for PMMA in the more restricted velocity range investigated
in \cite{Pauline2}. These authors have then been able to deduce values of
the volume $v_{act}$. In both cases, they find it to be on the order of
a few nm$^{3}$.

This naturally leads to conclude that, inspite of its crudeness, the
toy model does capture the main physical mechanisms responsible for
the frictional rheology of MCI, namely~:

    {\it (i)} When sheared, the highly confined junctions between
    asperities are the seat of mechanical instabilities, each of which
    primarily affects a small region containing a few atomic (or, for
    polymer glasses, monomer) units.

    {\it (ii)} The strengthening, logarithmic, rate effect can be
    assigned to premature ``flips" of these clusters induced by {\it
    thermal}
    noise.

Feature {\it (i))} is fully consistent with our description of the
junction as a quasi-$2D$ disordered medium, solidified under the
high confinement conditions imposed by the bulks of the contacting
asperities, but
mechanically weaker than these bulks -- probably due to a higher
fraction of quenched free volume. Indeed, a number of numerical
studies, pioneered by Argon et al \cite{Argon} (see also Falk and
Langer \cite{Falk} and Malandro and Lacks \cite{Lacks}) of sheared glasses,
both
polymeric and molecular, at temperatures $\ll T_{g}$, have shown that
elastic dissipation in these systems occurs essentially via sudden
collective rearrangements of clusters of typically nanometric volume.
These clusters, termed by Falk and Langer shear transformation zones
(STZ), are randomly located
\footnote{ In the absence of shear localization}
and their average density is constant at constant
shear rate $\dot{\gamma}$
\footnote{To which extent and how this density depends on $\dot{\gamma}$
is an important though still unsettled
question.}
: on average, once a STZ has flipped into a lower energy local
equilibrium state, another one appears at an uncorrelated position.
That is, we may interpret the interblob bistable contacts of the toy
model as a sketchy representation of STZ's, the volume $v_{act}$
providing a rough evaluation of the average zone volume.

Feature {\it (ii)} opens onto a much more difficult, still essentially
unsolved question. In the toy model, the bistable units are completely
decoupled, and their contributions to the friction stress are simply
additive. The real junction is, of course, dense everywhere, so that
each STZ must be understood as embedded in a quasi-$2D$ elastic medium,
and coupled to the adjacent deformable asperity bulks which, though
stiffer, do transmit stresses. The flip of a STZ is therefore akin to
the tranformation of an Eshelby inclusion \cite{Eshelby}: as already
emphasized in
\cite{Argon}, it gives rise to a multipolar force signal which
deforms the surrounding medium \cite{Ajdari}, hence resulting
in loading steps on the other STZ's, which were ``on their way towards
flipping". The randomness of these signals in time and space is the source
of the so-called {\it dynamic noise}, which acts in parallel with the
thermal one. Elastic coupling being long ranged, dynamic noise is
likely to trigger cascades (avalanches) of correlated flips, which
are not accounted for in mean field approximations -- hence the
difficulty of this class of problems \cite{Fisher}.

Moreover, clearly, the larger the imposed strain rate $\dot{\gamma}$,
the more frequent the flips hence, roughly speaking, the larger the
effective strength of the dynamic noise -- which must vanish in the
vanishing $\dot{\gamma}$ limit. So, if these effects can be modelled
in terms of an effective temperature, this must be a growing function
of $\dot{\gamma}$. Up to now, no theory of this effect is available,
and it is therefore not taken into account by the recent
phenomenological theories of  plasticity of amorphous media~: the
STZ-based one \cite{Falk} and the soft glass rheology of Sollich et
al. \cite{Sollich}. Qualitatively, one expects such rheologies to
exhibit a crossover between two limiting regimes~: a thermal
noise-controlled one at low $\dot{\gamma}$ and/or high $T$, a dynamic
noise-controlled one in the opposite conditions.

Our above analysis of friction leads us to conclude that {\it the rheology
of MCI junctions is controlled , at and above room
temperature, by thermal noise, in the investigated range of
$\dot{\gamma} = V/h$ i.e. typically $\dot{\gamma} \lesssim 10^{5}$}.

\begin{figure}[h]
     \includegraphics[width = 7cm]{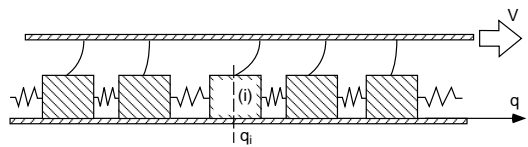}
    \caption{
    \label{fig:bo}
    Schematic representation of Persson's model \protect\cite{Bo}. The
    junction is constituted of a dense set of elastically coupled
    nanoblocks driven by a rigid plane through shear springs. When the
    stress on a block $\sigma < \sigma_{a}$, it is stuck to the lower
    plate. When $\sigma$ reaches $\sigma_{a}$ it gets depinned and
    performs damped motion. When its sliding velocity $\dot{q}_{i}$
    vanishes it resticks.}
    \end{figure}

Persson has studied numerically a model which represents the junction,
as sketched on Figure (\ref{fig:bo}), as a dense set of elastically
coupled pinned blocks driven by a rigid plate. He found  (see
\cite{Bo}, chap. 11) that dynamic noise effects were negligibly
small, thus justifying a thermal noise-controlled mean field rheology
(equation (\ref{eq:fric-activ})). Note however that the smallness of
dynamic noise effects in his model is likely to result from the
infinite stiffness of the confining plates, which should lead to
exponential screening of elastic couplings.\\

\begin{figure}[h]
     \includegraphics[width = 7cm]{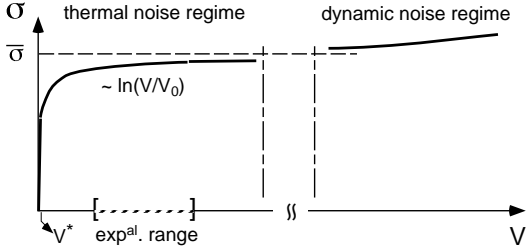}
    \caption{
    \label{fig:artist}
    Schematic interfacial stress {\it vs} velocity curve (see text). The
    Newtonian Eyring regime corresponds to $0 < V < V^{*}$.}
    \end{figure}

Our analysis of MCI rheology can therefore be summed up by the
artist's view of Figure (\ref{fig:artist}). The
"threshold-plus-logarithmic" behavior fitted by the toy model
results should therefore be understood as corresponding to an
intermediate regime, with limits on both the high and low velocity sides.

In view of the current high activity in the field of soft glass
rheology, one may be optimistic about the emergence of theoretical
predictions about the position of the high-$V$ crossover towards the
non thermal regime. However, it must be pointed out that
MCI friction is certainly not a good tool for experimental
investigation of this question since, at the necessary sliding
velocities, in the mm/sec range or larger, one would have to cope with
the tricky problems related with e.g. frictional heating.

The opposite low-$V$ limit is essentially academic. Let us come back
to the toy model~: as the stress decreases, the cut-off $X_{c}$ of
the contact distribution approaches form above the position
corresponding to the Maxwell plateau, and the rate of back jumps from branch
$2$ to $1$ becomes non negligible and comparable with the $1
\rightarrow 2$ one. At the same time, the interbranch barrier
$\Delta{\cal E}_{b}$ increases to $\sim \sigma_{Y}v_{act}$, making
jumps extremely rare. This results in the classical Eyring ($sinh$)
behavior for $\sigma_{s}(V)$, ending in an extremely fast
quasi-linear drop to zero, with slope $\eta/h$, where $\eta$ is the
viscosity of the glassy junction. The crossover velocity $V^{*}$ is
in practice much to small to be observable.

\subsection{Sliding dynamics of a MCI~:}

\subsubsection{The Rice-Ruina friction law~:}

We can now sum up the above phenomenological analysis into a
model of MCI dynamical friction which reads, with the help of equations
(\ref{eq:age4}), (\ref{eq:age9}),(\ref{eq:age10}) and
((\ref{eq:sigmas})~:

\begin{equation}
\label{eq:RR1}
F(\phi, \dot{x}) = \sigma_{s0} \Sigma_{r0}\,\left[1 + m \ln \left(1 +
\frac{\phi}{\tau}\right) \right]\,\left[1 + \alpha \ln \left(
\frac{\dot{x}}{V_{0}}\right) \right]
\end{equation}

\begin{equation}
\label{eq:RR2}
\dot{\phi} = 1 - \frac{\dot{x} \phi}{D_{0}}
\end{equation}
We saw that both $m$ and $\alpha$ are typically of order $10^{-2}$,
so that it appears reasonable to neglect the $m\alpha \ln^{2}$ term in
expression (\ref{eq:RR1}) for the friction force. Moreover, we saw
that the short time cutoff $\tau$ associated with the early stage of
area creep becomes relevant only for velocities $\sim D_{0}/\tau$, on
the order of a few hundred $\mu$m/sec at least. For the moderately
accelerated instationary dynamics which we will analyze below, in
order of magnitude $\phi \sim D_{0}/\dot{x}$ so that, in the low
velocity regime considered here, $\phi/\tau \ll 1$. Equation
(\ref{eq:RR1}) then reduces to the following expression for the
dynamic friction coefficient~:

\begin{equation}
\label{eq:RR3}
\mu_{d}\left(\phi, \dot{x}\right) = \mu_{d}(V_{0}) + B
\ln\left(\frac{\phi V_{0}}{D_{0}}\right) +
A\ln\left(\frac{\dot{x}}{V_{0}}\right)
\end{equation}
where the reference velocity $V_{0}$ can be chosen at
will in the above-mentioned velocity range. The constants $B$ and $A$
are defined in equations (\ref{eq:age6}) and (\ref{eq:alpha}).

{\it Equations (\ref{eq:RR2}) and (\ref{eq:RR3}) constitute the Rice-Ruina
(RR) model, proposed by these authors in 1983 on the basis of Dieterich's
experiments.}

\subsubsection{The RR dynamics of a driven block~:}

The question then immediately arises of analyzing and testing
experimentally the predictions of the model in terms of sliding
dynamics. First of all, does it correctly describe the stick-slip (SS)
oscillations of a driven block, and their disappearance at high
enough stiffness?

Consider the system depicted on Figure \ref{fig:schema}. Let us assume
for the moment that the sliding velocity is uniform along the
interface (homogeneous sliding), i.e. that the block has a single
degree of freedom, the position $x(t)$ of e.g. its center of mass. Its
equation of motion reads~:

\begin{equation}
\label{eqRR4}
M \ddot{x} = - K \left(x - x_{0}(t)\right) - W \mu_{d}(\phi, \dot{x})
\end{equation}
where $\mu_{d}$ and $\phi$ are specified by equations (\ref{eq:RR3})
and (\ref{eq:RR2}), and $(x-x_{0}(t))$ is the instantaneous elongation
of the driving spring.

At any pulling velocity $V$ there always exists a steady sliding
solution, namely~:

\begin{equation}
\label{eq:RR5}
\dot{x} = V \,\,\,\,\,\,\,\,\, \phi = \frac{D_{0}}{V}
\,\,\,\,\,\,\,\,\,
x(t)-x_{0}(t) =  \frac{W}{K} \mu_{d}(\frac{D_{0}}{V}, V)
\end{equation}

In view of the non-linearities of the friction law, one must wonder
about its dynamic stability, i.e. perform a standard linear stability
analysis~: setting $\dot{x} = V + \delta\dot{x}(t)$, $\phi =
D_{0}/V + \delta\phi(t)$, one linearizes the dynamical equations in
$\delta\dot{x}$, $\delta\phi$. Thanks to the time invariance of the
basic state, the solutions for them are of the form $Const \exp(i\Omega
t)$. Steady sliding is stable (resp. unstable) when $Im\,\Omega > 0$
(resp. $<0$).

This calculation is performed in Appendix C. It shows that, for given
values of $M$, $W$ and $V$, steady motion is stable for $K > K_{c}$,
with the critical stiffness given by:

\begin{equation}
\label{eq:Kc}
K_{c} \frac{D_{0}}{W} = \left(\mu_{\phi} - \mu_{\dot{x}}\right)
\left[ 1 + \frac{MV^{2}}{WD_{0}\mu_{\dot{x}}}\right]
\end{equation}
At the bifurcation point ($K = K_{c}$) $\Omega$ is pure real and has
the value :

\begin{equation}
\label{eq:omegac}
\Omega_{c} = \frac{V}{D_{0}} \sqrt{\frac{\mu_{\phi} -
\mu_{\dot{x}}}{\mu_{\dot{x}}}}
\end{equation}

That is, the corresponding bifurcation is of the Hopf type, signalling
that for $K \leq K_{c}$ motion should become oscillatory.

In equations (\ref{eq:Kc}), (\ref{eq:omegac}), $\mu_{\phi}
= \partial\mu_{d}/\partial(\ln\phi)$,
$\mu_{\dot{x}} = \partial\mu_{d}/\partial(\ln\dot{x})
V$, both derivatives being evaluated at $\phi = D_{0}/V,\,\dot{x} = V$.
The RR
expression results in $\mu_{\phi} = B$, $\mu_{\dot{x}} = A$. In the
experiments aiming at characterizing the SS bifurcation, the block
was sliding under its own weight~: $W = Mg$. Then with $D_{0} \sim 1
\mu$m$/sec$, $A \sim 10^{-2}$, the inertial correction in the second
factor of the r.h.s. of equation (\ref{eq:Kc}) is of order
$10^{-5}(V_{\mu m/s})^{2}$. So, inertia is negligible for $V \lesssim
100\mu$m$/sec$. Then the RR model predicts that the critical stiffness

\begin{equation}
\label{eq:KcRR}
K_{c} = \left(B -A\right)\,\frac{W}{D_{0}}
\end{equation}
should be velocity-independent, while the critical pulsation

\begin{equation}
\label{eq:omegacRR}
\Omega_{c} = \frac{V}{D_{0}}\,\sqrt{\frac{B - A}{A}}
\end{equation}

\begin{figure}[h]
     \includegraphics[width = 7cm]{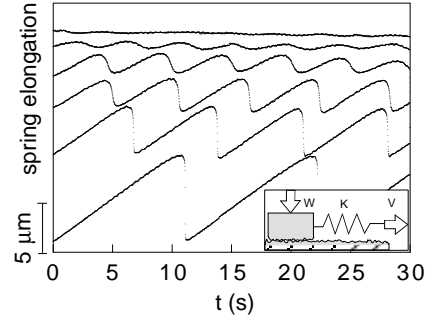}
    \caption{
    \label{fig:papier}
    Spring elongation (see inset) versus time for a
    paper/paper
    MCI pulled at the fixed velocity $V = 1 \mu$m/sec. As $K/W$
    decreases, sliding motion bifurcates continuously from steady (upper
    curve) to periodic stick-slip of growing amplitude.}
    \end{figure}

\begin{figure}[h]
     \includegraphics[width = 7cm]{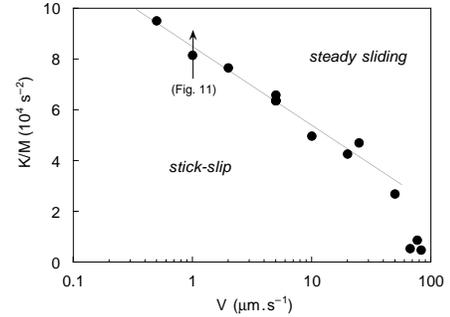}
     \caption{
    \label{fig:bifPMMA}
    Dynamic stability diagram of a PMMA/PMMA
      MCI in the low velocity range. Steady sliding is stable above the
      bifurcation curve $(K/W)_{c}(V)$.}
  \end{figure}

The SS bifurcation has been studied on symmetric MCI involving paper
\cite{Heslot}, glassy PMMA and PS \cite{Pauline2}. In these
experiments, $K$ was kept constant. Then, when $M$ was increased at
constant $V$, the initially steady motion was observed to bifurcate
at a value $(K/M)_{c}$ beyond which it develops oscillations whose
amplitude grows continuously as $K/M$ decreases, until true sticking
phases appear (Figure \ref{fig:papier}). Tracking the bifurcation
at various $V$'s leads to the stability diagram in the plane ($K/M, V$) of
the control parameters, a typical example of which is displayed on
Figure \ref{fig:bifPMMA}. It is seen that $(K/W)_{c}$ is not strictly
constant, but decreases slowly with increasing $V$ at a rate on the
order of $10$ per cent per decade. On the other hand, $V/\Omega_{c}$ was
measured to be constant up to experimental accuracy
\footnote{ The poor accuracy on $\Omega_{c}$ was due in particular to
the difficulty of extrapolating measurements necessarily performed at
finite amplitude to the continuous bifurcation while, in this case,
non linearities develop very fast \cite{HopfNL}.} \cite{Heslot}.
So, the predictions of the RR model appear qualitatively satisfactory.
    
The parameters $A, B, D_{0}$ of the model are then evaluated as
follows~: measurements of $d\mu_{d}/d(\ln V)$ yield $(B - A)$, $D_{0}$
is obtained by equation (\ref{eq:KcRR}) from the value of $(K/W)_{c}$
at some reference velocity -- e.g. $1 \mu$m$/sec$ -- then $A$ is
determined with the help of equation (\ref{eq:omegacRR}). For example,
for a PMMA MCI, one thus obtains \cite{Cochard}~: $A = 1.2\,10^{-2}
\pm 2\,10^{-3}, B = 2.3\,10^{-2} \pm 2\,10^{-3}, D_{0} = 0.4 \pm 0.04
\mu$m. Such parameter values are fully compatible with those deduced
from static aging and velocity jump experiments.

In order to test more thoroughly the validity of the phenomenological
model, we have analyzed whether it correctly predicts a few other
salient features of MCI block dynamics, among which~:\\

(i) {\bf Non-linear development of SS-like oscillations close to
the bifurcation:} it must be mentioned first that the RR model as
such (constant $\mu_{\phi} = B$ and $\mu_{\dot{x}} = A$) results in a
highly singular behavior for $K \leq K_{c}$ \cite{Ranjith}, namely
unbounded growth of the velocity at finite time. This unphysical behavior, in
contradiction with the observation of a continuous bifurcation, can
be traced back to the simplifying assumption that $\mu_{\phi}$ and
$\mu_{\dot{x}}$
are mere constants, which is also responsible for the fact that the model
fails to
accounts for the non zero slope of the $K_{c}(V)$ bifurcation curve.
Baumberger et al \cite{HopfNL} relaxed this assumption by assigning the
slope of $K_{c}(V)$ to a $(\ln^{2}\phi)$ correction to
expression (\ref{eq:RR3}) for the friction coefficient. A standard
perturbation expansion then results
\footnote{The expansion parameter is found to be $\epsilon \sim
A/\mu_{2}$ with $\mu_{2} = [\partial^{2}\mu_{d}/\partial(\ln x)^{2}]$,
so that for the RR model ($\mu_{2} = 0$) the expansion explodes. For
PMMA, $\mu_{2}$ is measured to be $\sim 10^{-3}$, hence $\epsilon \sim 10$
: non
linearities develop fast in the SS regime.}. This extension of the RR
constitutive law accounted very satisfactorily for the growth of the
oscillation amplitude and of the frequency shift with $(K_{c} - K)$.\\

\begin{figure}[h]
     \includegraphics[width = 7cm]{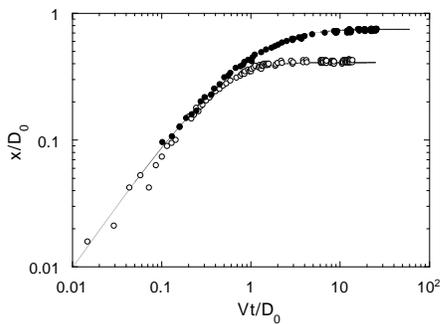}
    \caption{
    \label{fig:relax}
    Log-Log plot of reduced slip distance $x/D_{0}$ {\it vs} reduced
    time $Vt/D_{0}$ during creep-like relaxation of a paper/paper MCI.
    The system was sliding at $V = 1 \mu$m/sec (open circles) and $10
    \mu$m/sec (full circles) until the drive was stopped at $t = 0$. The
    lines are the fits obtained from the RR model with a common set of
    parameters. }
    \end{figure}

(ii) {\bf Creep-like sliding motion following cessation of the
drive:} As can be seen on Figure \ref{fig:stop and go MCI} once the
drive at velocity $V$ is stopped, the
block continues to slide while slowing down until it stops at a
finite force level. Baumberger and Gauthier \cite{TBrelax} analyzed
the dependence of such relaxation curves on the control parameters
$K/W$, $V$. They showed that the RR model predicts that the total
distance slipped before arrest should be $V$-independent, while
experiments show a $100 \%$ increase on one decade of $V$. Again,
this discrepancy is cured by including into equation (\ref{eq:RR3})
the same $\ln^{2}\phi$ correction as for non linear effects close to
the bifurcation. An example of the resulting fit of relaxation data is
shown on Figure \ref{fig:relax}.\\

These two examples indicate the limits to be assigned to the rate and state
model in its simplest version. While, clearly, the RR law does
capture the essential features of MCI friction -- namely geometric
aging and dissipation governed by thermally-assisted mechanical
instabilities on the nanometer scale -- it is insufficient to account
for finer dynamical features involving finite departures from
stationary sliding. As described above, phenomenological corrections
to the expression of $\mu_{d}$ permit to extend the validity of the
model. However, such extensions must be regarded with proper caution~:
clearly, the non-zero slope of the $K_{c}(V)$ bifurcation curve
indicates that an extension of expression (\ref{eq:RR3}) is
needed. But this leaves open the question of the respective weights
of the various possible corrections which we have mentioned when
deriving expression (\ref{eq:RR3}). Not only is a $\ln^{2}\phi$ term
possible, but also corrections of the form $(\ln\dot{x})^{2}$
and $(\ln\phi)(\ln\dot{x})$ as well as higher order ones. Moreover,
corrections associated with the short age time cut-off $\tau$ might
become relevant. This would mean introducing many more parameters,
in principle to be fitted from strongly non-linear dynamical features.
As shown by the detailed analysis of the response to to large amplitude normal
force oscillations \cite{oscnorm} \cite{Cochard}, such fits become of
more doubtful value as the dynamical complexity increases. This
points towards the fact that formal extensions, or regularizations, of
such a constitutive law, though they may appear useful, are in general
difficult to legitimate in detail on a physical basis. So they must be
considered with a critical eye when using them for predictive purposes.

\subsubsection{Limitations of block models for extended MCI~:}

Seismic faults may be represented as spatially extended MCI under
both normal and tangential loading. Studying their dynamics then
means solving for the elastodynamic motion of two ``semi-infinite"
deformable media in frictional contact. It was the need for a realistic
constitutive law for fault friction which motivated the work which
resulted in the formulation of the RR model.

Once such a law is available, one is naturally led to plugging it into a
continuum mechanics description. On the other hand, note that, since seismic
events amount to the nucleation and propagation of interfacial
cracks, they involve variations of deformation fields down to the very
small scales relevant close to crack tips.

This leads us to an
important remark resulting from the above analysis of the physics
involved in the RR law. We saw that this law emerges from two
averaging processes~:

  (i) the rheology described by the interfacial strength $\sigma_{s}$
  results from averaging over dissipative events involving
  nanometric regions within junctions; its expression is thus valid
  only on scales much above the nanometer one.

  (ii) the creep growth of the real area of contact is translated into
  an geometric age  which is an average over a large number of
  microcontacts. It thus only makes sense on a scale much larger than
  intercontact distances, i.e. typically millimetric.

As already mentioned, this makes the much discussed question of the
ill-posedness
\cite{Ranjith-Rice} of the
continuous limit a formal one. In view of the above
remarks, which lead to the existence of a physically based small space
scale cut-off, the most physical recipee for curing mathematical
ill-posedness seems to be the one proposed by Simoes and Martins
\cite{Simoes}~: they wash out the UV singularity by replacing the
local friction relation between interfacial normal and tangential
stresses $ \left(\sigma_{t}(x) = - [\mu_{d} \sigma_{n}]_{x}\right)$ by a non
local one $ \left(\sigma_{t}(x) = - \int dx'w(x - x')[\mu_{d}
\sigma_{n}]_{x'}\right)$, with $w$ a spreading function of finite
width.\\

In practice, the dynamic complexity due to the non-linearities of MCI friction
\cite{Carlson-Langer} can only be studied numerically. This is
implemented by discretizing the system into elastic blocks of size
$L\times L\times L$
\footnote{Choosing blocks much longer in the transverse than in the
longitudinal direction (i.e., in Burridge-Knopoff-like models,
compressive
stiffnesses much larger than shear ones) may lead to dynamical artefacts~:
for the so-called small events which involve only a few blocks, the
associated elastic fields only affect depths, on the order of
their lateral extent, much smaller than the block height. It is then
illegitimate to neglect internal block degrees of freedom.}.
Rice\cite{RiceBK} has pointed out that the existence of the stick-slip
instability imposes a natural upper limit on this block size. Indeed,
such a discretization implies that the degrees of freedom of a block reduce to
those of, say, its center of mass, internal ones being irrelevant. A
block of size $L$ has a tangential stiffness $K_{L} \sim EL$, with $E$
an elastic modulus of the bulk. It bears the normal load $W_{L} =
\sigma_{n}L^{2}$, with $\sigma_{n}$ the average normal far field
stress. It is stable against SS if $K_{L}D_{0}/W_{L} > (B -A)$, i.e.
$L < L_{R}$ with~:

\begin{equation}
\label{eq:LRice}
L_{R} = \frac{E}{\sigma_{n}}\,\frac{1}{B -A}\,D_{0}
\end{equation}

Assume that a block has a size $L_{R} < L < 2L_{R}$. Cut it into $8$
subblocks of size $L/2$. Each interfacial subblock is stable
vis-\`a-vis SS, while the block is not. So the whole dynamic
complexity on scale $L$ is associated with the motion of the relative
subblock position, which cannot therefore be considered irrelevant.

The maximum size of a discretization block is therefore $L_{R}$. From
equation (\ref{eq:LRice}), with $\sigma_{n}/E \sim 10^{-3}$, $B-A \sim
10^{-2}$, $ D_{0} \sim 1$--$ 10 \mu$m, $L_{R}$ lies in the meter range.\\

\section{JUNCTION RHEOLOGY~: STRUCTURAL AGING/REJUVENATION EFFECTS}

It is clear, at this stage, that the rough-on-rough MCI configuration
only provides quite an indirect access to the analysis of junction
rheology. Indeed, information about the interfacial strength
$\sigma_{s}$ can be extracted from friction data only after
"deconvoluting" it from geometric aging effects, necessarily at the
expense of precision. This might lead to overlooking finer physical
features of dissipation within sheared junctions.

That such might indeed be the case emerged from the experimental
study by Bureau et al \cite{jamming} of the response of a rough/rough
PMMA MCI to an oscillating shear force~:

\begin{equation}
\label{eq: jam}
F(t) = F_{0} + f \cos \omega t
\end{equation}
biased about a value $F_{0}$ below the static threshold. For very
small $f$, the MCI responds elastically. As $f$ is increased, the
system enters a regime in which the oscillatory response is superimposed
on a slow, self-decelerating, gross sliding motion which corresponds
to a saturating, finite displacement. This regime prevails in a narrow
amplitude range where $F_{max} = F_{0} + f \gtrsim F_{s}$, where $F_{s}$ is
a static
threshold. It was analyzed, on the
basis of the RR model, as resulting essentially from the geometric age
dynamics. During each force oscillation, the system alternates
between sliding, hence rejuvenating, when $F(t)$ is close below its
maximum, and slowing down strongly, thus aging during the rest of the
period. For $f$ levels such that aging barely wins, gross motion
decelerates and the system finally jams. Above a threshold $f_{>}$,
rejuvenation wins, resulting in indefinite accelerated sliding.

The RR model accounts satisfactorily for the long time dynamics in the
jamming regime. However, it exhibits, for small
slid distances on the order of a few $100$ nm, a clear discrepancy with
  experimental results which strongly suggests
that another rejuvenation mechanism, distinct from the geometric one,
hence ignored by the RR description, might be at work within the junctions.

\subsection{ Accessing junction rheology directly~: suitable
configurations}

This confirms the interest of studying junction rheology directly,
with the help of interfacial configurations which either are free of
geometric aging or permit to circumvent this effect. This can be
achieved in three different ways.

  \subsubsection{Rough-on-flat multicontact interfaces}

  So far, we have restricted the definition of MCI to interfaces of
  macroscopic extent between two rough surfaces. It is then tempting,
  in order to get direct information about junction rheology, to try and
  realize interfaces {\it \ˆ la} Greenwood which would not undergo
  the microcontact birth and death process responsible, for
  rough-on-rough systems, for the non trivial dependence of geometric age on
  the sliding dynamical history.

  Such a configuration has been realized by Bureau et al \cite{Lionel EPJB}
   who studied friction between a rough PMMA slider (roughness
  $\sim \mu$m) and a flat and smooth plate made of float glass
  (roughness $\sim$ nm)
 ~\footnote{ Since glass is much harder than PMMA, no ploughing of the
  track by the slider asperities takes place.}.

\begin{figure}[h]
    \includegraphics[width = 7cm]{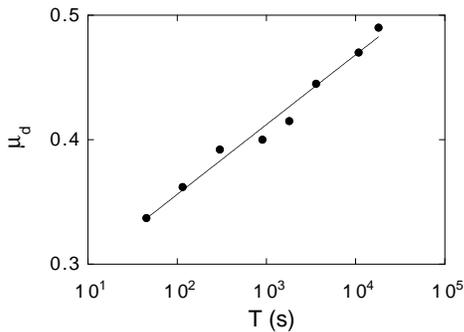}
    \caption{
    \label{fig:driftmud}
    Logarithmic growth of the {\it steady sliding} friction coefficient
    $\mu_{d}$ at velocity $V = 50 \mu$m/sec for a rough PMMA/flat glass
    MCI {\it vs} time $T$ elapsed since the interface was created.
    Adapted from ref. \protect\cite{Lionel EPJB}.}
    \end{figure}

  The interface is of the GW type but, due to translational invariance
  of the track, {\it the integrity of the contact population is now
  preserved upon sliding}. The load-bearing asperities creep as in a
  regular MCI but, since aging is not interrupted by sliding, the
  geometric age is simply the time $T$ elapsed from the creation of the
  interface. The real area of contact $\Sigma_{r}$ grows logarithmically
  with $T$ so that, in steady motion at velocity $V$, the friction force
  $F = \Sigma_{r}(\phi = T).\sigma_{s}$(V), becomes an increasing
  function of $T$, as can be seen on Figure \ref{fig:driftmud}.
  One can then take advantage of the slowing down of the logarithmic
  growth mode by letting the interface ''mature'' by waiting up to
  large $T$'s (typically $10$ hours), then performing experiments during
  a comparatively short period (e.g. $\lesssim 1$hour) during which
  $\Sigma_{r}$ remains quasi-constant, so that friction force
  variations are directly attributable to those of the interfacial
  strength $\sigma_{s}$.
  
  \subsubsection{Junctions in the Surface Force Apparatus}

\begin{figure}[h]                                                             
    \includegraphics[width = 5cm]{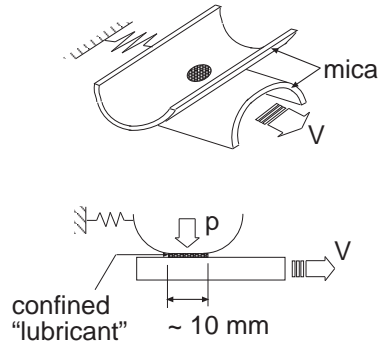}                             
    \caption{                                                                   
    \label{fig:SFA}                                                             
    The cross-cylinder geometry used in the SFA (upper sketch) is               
    equivalent to a boundary lubricated sphere-plane contact (lower sketch).}   
    \end{figure}

  The surface force apparatus (SFA) was first developed
  \cite{Tabor-Winterton} \cite{IsraelSFA} to study adhesion. In a SFA,
  two atomically planar mica surfaces, slightly curved so as to realize
  a cross-cylinder geometry (Figure \ref{fig:SFA}), are brought
  into contact in the
  presence of a fluid (the lubricant). The interplate distance is
  decreased down to nanometric values on the order of a few molecular
  sizes. In this {\it boundary lubrication} regime where the fluid  is
  very highly confined, the cylinders deform elastically, leading to a
  Hertz planar circular contact, with radius commonly in the $10  \mu$m
  range.

  The SFA was later modified \cite{JacobSSR} so as to allow for shear
  motion and measurement of friction forces. Since the area is measured
  optically, the normal stress borne by the contact is directly
  accessible, as well as the shear stress associated with frictional
  dissipation in the {\it junction formed by the lubricant}. In many
  instances (see below) such boundary lubrication layers behave as weak
  disordered solids, exhibiting an elastic regime at low shear levels.

  \subsubsection{Extended soft contacts}

  We refer here to the contacts which form between a highly compliant
  (soft) slider with a smooth surface and a hard flat smooth track. High
  slider compliance ensures that molecular adhesive contact is realized
  everywhere along the interface, even in the presence of submillimetric
  roughness.

  Such contacts can be achieved in either of two configurations~:

     --- {\it flat-on-flat}: up to now, this geometry has been used with
  sliders made of moulded hydrogels (Young moduli in the $1$ -- $10$ kPa
  range) \cite {Gong} \cite {gelEPJ}. The contact lateral extent is
  usually of order centimeters.

     --- {\it ball-on-flat}: this is the configuration commonly used
  to study friction at elastomer/glass interfaces \cite{Schall}
  \cite{Barquins} \cite{Chaudhury}. Pressing the ball onto the flat
  under controlled normal load results in the formation of a circular
  adhesive Hertz contact \cite{Johnson}. Since elastomer
  Young moduli typically lie in the $1$ -- $10$ MPa range, contact
  radii in such experiments are commonly of order millimeters.

  \subsection{MCI junctions revisited~:}

  The results described here were obtained by Bureau et al
  \cite{Lionel EPJB} on MCI formed by rough PMMA sliding on smooth
  (float) glass. The surface state of the glass was controlled by
  grafting onto it a single monolayer of short silane molecules, which
  passivate strongly adhesive sites.

  As already mentioned, such interfaces, when ``old enough", can be
  considered to work at constant area of contact $\Sigma_{r}$, thus giving
direct
  access to variations of the frictional stress
 ~\footnote{ As the absolute value of $\Sigma_{r}$ is not measurable
  accurately, the absolute value of $\sigma_{s}$ cannot be accessed in
  this configuration.}.

  \subsubsection{Structural aging~:}

\begin{figure}[h]
    \includegraphics[width = 7cm]{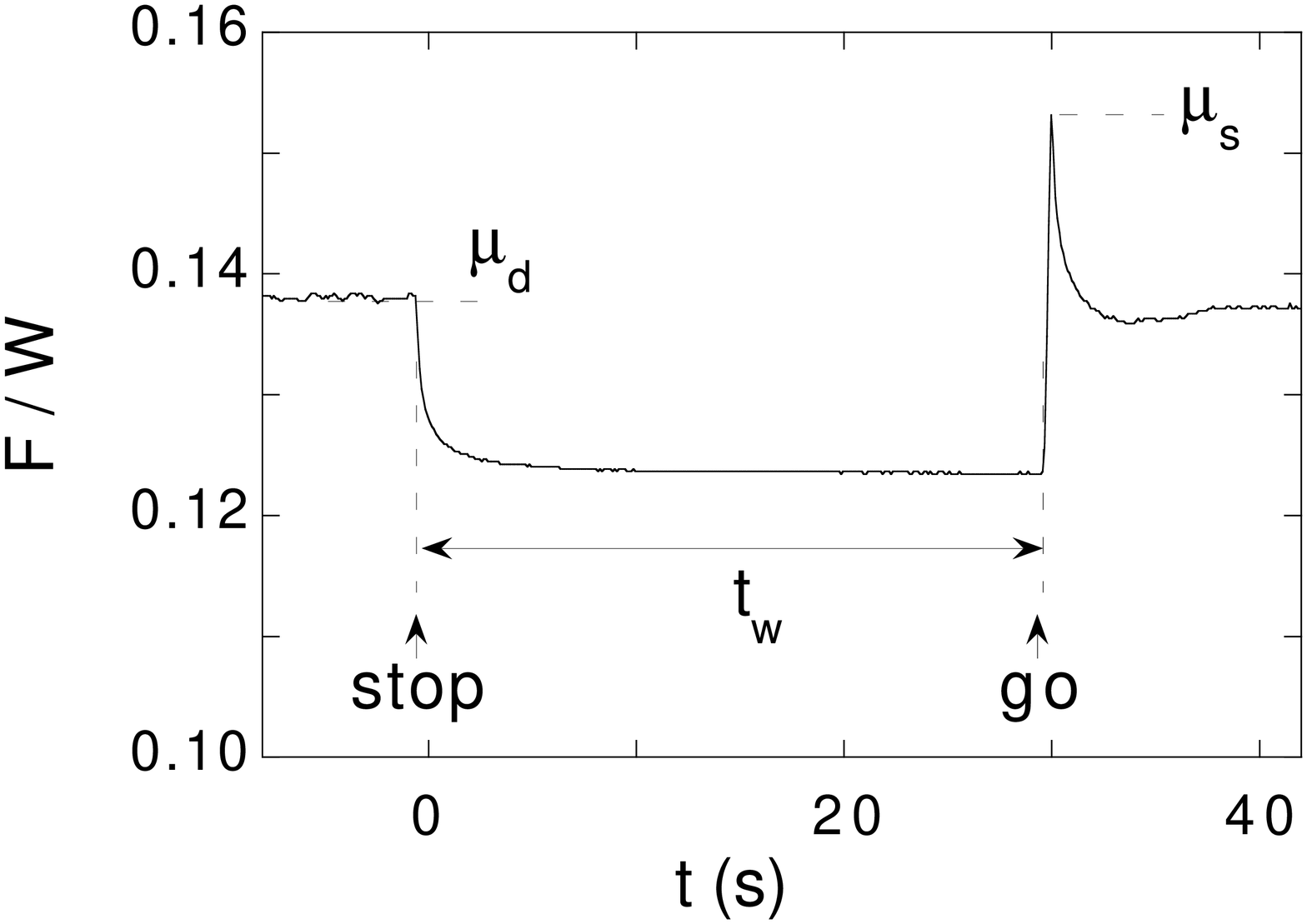}
      \caption{
    \label{fig:stop-go-lisse}
    Reduced shear force on a rough PMMA/silanized flat glass MCI in a
    stop and go experiment. For  $t < 0$ the block slids steadily ($V =
    10 \mu$m/sec). After the drive is stopped ($t = 0$) the slider slows
    down and stops at a self-selected stress level. At $t = 30$ s loading
    is resumed at the initial velocity.
    Adapted from \protect\cite{Lionel EPJB}.}
    \end{figure}

\begin{figure}[h]
    \includegraphics[width = 7cm]{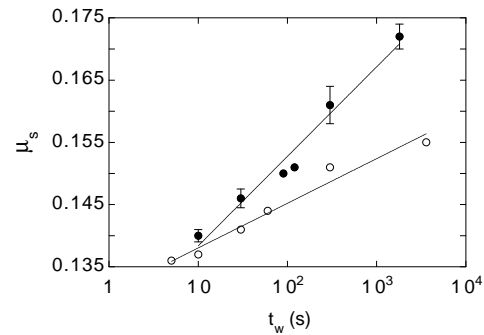}
    \caption{
    \label{fig:mus(t)-lisse}
    Logarithmic growth of the static friction coefficient $\mu_{s}$ with
    waiting time $t_{w}$. Aging at rest under~: ($\bullet$) the
    self-selected arrest shear stress; (o) null stress. From reference
    \protect\cite{Lionel EPJB}.}
    \end{figure}

  The central result is illustrated on Figure \ref{fig:stop-go-lisse}.
  Namely, standard stop-and-go experiments (see $\S$II.B.1) reveal the
  presence of a static friction peak, the amplitude of which grows
  logarithmically with the waiting time $t_{w}$ (Figure
  \ref{fig:mus(t)-lisse}).

  Contrary to what is the case for rough/rough MCI,
  this peak cannot be associated  with geometric aging, which is not
  operative here. On the other hand, the very existence of a
  $\mu_{s}(t_{w})$ larger than the dynamic $\mu_{d}$ means that the
interface is the seat of an aging-when-waiting {\it vs}
rejuvenating-when-sliding process, which must therefore necessarily take place
within the junctions and affect their structure. The existence of
this {\it structural aging} entails that $\sigma_{s}$ is not, as we
assumed up to now, a mere function of the instantaneous sliding
velocity $\dot{x}$, but does itself depend upon the dynamical history
of the interface.

Moreover, one observes that the structural aging rate depends
noticeably on the value of the tangential stress applied during the
stop phase (see Figure \ref{fig:mus(t)-lisse})~: the lower its level, the
slower aging is (see figure 5 of reference \cite{Lionel EPJB}).

Finally, in many cases -- and especially for small waiting
stresses -- structural aging becomes observable only after a finite,
strongly sample-dependent, {\it latency time} $\tau_{L}$.

\subsubsection{ Steady sliding dynamic friction~:}

\begin{figure}[h]
    \includegraphics[width = 7cm]{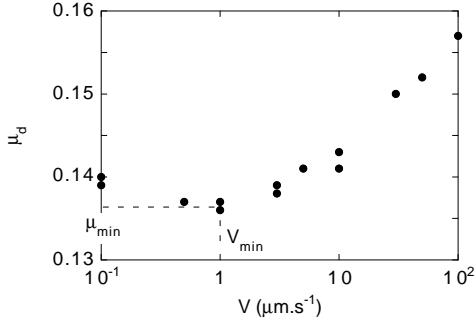}
    \caption{
    \label{fig:mud-lisse}
    Dynamic friction coefficient $\mu_{d}$ {\it vs} velocity $V$.
    From reference \protect\cite{Lionel EPJB}.}
    \end{figure}

Figure \ref{fig:mud-lisse} shows a typical example of the velocity
dependence observed for $\mu_{d}$. It exhibits a minimum at $V = V_{min}$.
The velocity-weakening observed for $V < V_{min}$ provides another
evidence that the structural age decreases with $V$, i.e. that sliding
induces rejuvenation.

For $V > V_{min}$, $\mu_{d}$ increases quasi-logarithmically
($\mu_{d}/\mu_{min} \approx (1 + \alpha \ln(V/V_{min}))$. That is,  we
recover here the expression of $\sigma_{s}$ (equation (\ref{eq:sigmas}))
deduced from the analysis of experiments on rough/rough MCI, which were
used to formulate the RR model. Moreover, the reduced slope $ \alpha
\simeq 4. 10^{-2}$ has a magnitude comparable with the value ($\alpha
\simeq 5.10^{-2}$) found for rough/rough PMMA junctions. This enables
us to conclude that already quite close above $V_{min}$, the
rejuvenation effect is quasi-saturated.

\subsubsection{Discussion}

This phenomenology, the fine details of which were masked, for
rough/rough sytems, by the larger effects of geometric aging
\footnote{As can be seen on Figure \ref{fig:mus(t)-lisse}, the
log-slope of $\mu_{s}$ associated with structural aging is on the order
of a few $10^{-3}$, typically from $10$ to $4$ times smaller than the
aging slopes for rough/rough MCI. This leads one to conclude that, in
the latter case, the growth of $\mu_{s}$, though dominated by the
geometric effect, does contain a small contribution due to structural
aging.}
, appears
qualitatively consistent with our description of junctions as
confined weak glassy media. Indeed, it is closely akin to the behavior
of sheared bulk soft glassy materials, such as colloidal glasses
\cite{Derec} \cite{Abou} and pastes \cite{Cloitre}. The rheology of
these systems has also been modelized \cite{Sollich} \cite{Fielding}
\cite{Lequeux} as resulting from the interplay between~:

  --- aging at rest, i.e. strengthening via slow relaxation down the
  highly multistable energy landscape characteristic of quenched
  disordered, jammed, media.

  --- rejuvenation by motion, which can be seen as reshuffling the
  populations of the local minima.

In the sliding (i.e. plastic) regime, the rate of the localized dissipative
events invoked in $\S$II.C.4 should be increased by the decrease of
the energy barriers due to stress induced biasing. The ``young",
shallow energy states which were depopulated as aging pushed the
system into ``older", deeper ones, thus can get repopulated~: the sheared
soft glass rejuvenates.

In the same perspective, one may tentatively interpret the
above-mentioned dependence of the aging rate on waiting stress level
as follows. In a stop-and-go experiment, the ``stop" acts as a
mechanical quench of the strongly rejuvenated formerly sliding state.
Aging then restarts in a landscape with barriers whose height is
decreased by stress-induced biasing, hence accelerates when the stress
level increases
\footnote{ Viasnoff and Lequeux \cite{Viasnoff} have found that
applying to a colloidal glass, aging under stress-free conditions, an
oscillating shear stress of finite duration does result in a strong
perturbation of the aging process. However, it is important to note
that, in their experiment, the stress amplitude is larger than the
yield stress of the material, which certainly results in some
non-stationary flow. This contrasts with the situation of ref.
\cite{Lionel EPJB},
where the applied waiting stress does not provoke any
sliding.}.

\subsection{ Boundary Lubrication Junctions}

Over the past twenty years a considerable body of results has come out of
friction experiments performed in the SFA, in the boundary lubrication
  (BL) regime. It is much too vast to be extensively reviewed here. We
  limit ourselves to those salient features which we consider to be most
  relevant in the perspective of this article.

  \subsubsection{Confinement-induced solidification}

\begin{figure}[h]
    \includegraphics[width = 5cm]{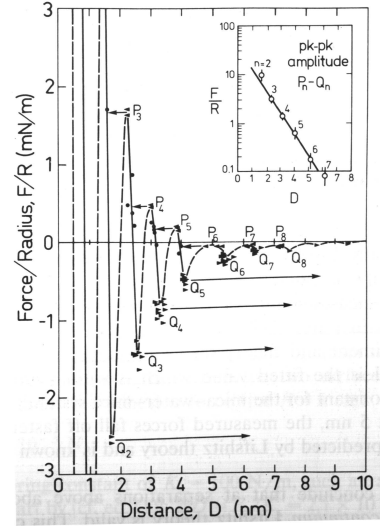}
  \caption{
    \label{fig:oscFn}
    Measured oscillatory force between two mica surfaces immersed in the
    liquid OMCTS. The arrows indicate inward or outward jumps from
    unstable to stable positions. Inset~: peak-to-peak amplitudes of the
    oscillations as a function of $D$.
    From ref. \protect\cite{JacobSSR}.}
    \end{figure}

  When the ``lubricant" -- a liquid in the bulk phase at the temperature
  of the experiment -- is compressed between the two mica surfaces, it
  is gradually squeezed out until its thickness is reduced to a few
  molecular sizes. Then, for materials composed of small, spherical or
  short-chain, molecules (e.g. OMCTS -- an inert silicone -- squalane, linear
  alcanes) one observes \cite {JacobSSR} that the normal force-{\it
vs}-thickness
  $F_{N}(D)$ curve develops oscillations whose amplitude grows rapidly
  as $D$ decreases (see Figure \ref{fig:oscFn}). This behavior can be
  ascribed to the layering of the confined fluid, each successive
  minimum being reached by squeezing out one more full layer.

  Thin enough (typically 3--4 molecular layers) such junctions, when
  sheared, exhibit an elastic response at low shear stress levels,
  until they reach a static threshold where they yield and start
  sliding \cite{Homola} \cite{Yoshi} \cite{van Alsten}. This
  demonstrates that they have solidified under the effect of the high
  confinement
 ~\footnote{ For some fluids made of short-chain molecules, e.g.
  hexadecane \cite{Yoshi}, the solid-like response only buils up after
  a finite amount of shear-induced sliding -- suggesting that in this
  case initial sliding helps ordering.}.
  This behavior has been confirmed by numerical studies \cite{Thompson}.

  Klein and Kumacheva \cite{Klein} found that, in the case of OMCTS
  (and at variance with previous results reported in \cite{van
  Alsten}), the liquid-to-solid transition occurs abruptly when
  decreasing the number of layers from $7$ to $6$. Whether or not such
  abruptness is a general feature has not been documented on different
  systems.

  Finally, it was shown in \cite{Yoshi} that increasing the
  temperature makes the confined solid weaker, i.e. leads to a
  decrease of its yield stress.

  It is important to note that the highest normal stresses which can
  be realized in the SFA are on the order of $10$ MPa. That is, they
  are at least one order of magnitude smaller than the levels met in
  MCI microcontacts, which we have seen to be comparable with the
  yield stress of the confining bulk solids.

  \subsubsection{Structural aging}

  Stop and go experiments show that the static threshold of
  solidified BL junctions increases with the waiting time $t_{w}$ spent
  at rest. This strengthening, which reveals structural aging, has
  been studied quantitatively on a variety of lubricants, including
  hexadecane \cite{Yoshi}, squalane \cite{Dru1} \cite{Gour} and a
  star-shaped polymer melt \cite{Yamada}. Strengthening also manifests
  itself through the increase with $t_{w}$ of the layer elastic
  stiffness \cite{Reiter}.

  \begin{figure}[h]
    \includegraphics[width = 5cm]{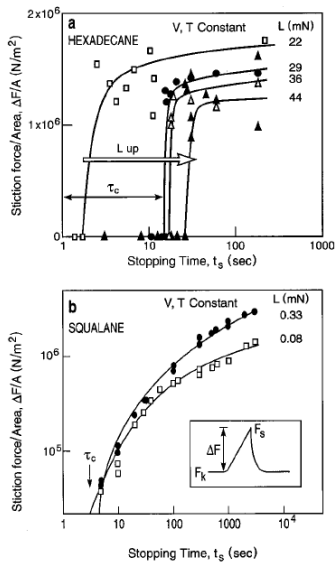}
  \caption{
    \label{fig:stiction SFA}
    Stiction peak heights {\it vs} waiting time $t_{s}$ for (a)
    hexadecane (from ref. \protect\cite{Yoshi}); (b) squalane (from ref.
    \protect\cite{Dru2}). $L$ is the normal load, $\tau_{c}$ the latency time.}
    \end{figure}

In all cases, strengthening is slow, namely either linear on a
  logarithmic $t_{w}$-scale
  \cite{Yoshi} \cite{Reiter} or somewhat faster \cite{Gour}.
As illustrated on figure \ref{fig:stiction SFA}, finite latency times,
ranging up to about $10$ seconds, previous to
  aging, have been observed on hexadecane \cite{Yoshi} and several
  lubricants including squalane \cite{Dru1}. This result was
  later contradicted, for squalane, in reference \cite{Gour} -- a
  discrepancy which is possibly ascribable to the difficulty of
  defining such a notion with high precision. Indeed, at short waiting
  times, the height of the stiction peak (the difference between the
  static threshold force and the dynamic one at the imposed driving
  velocity) becomes very small. Since accuracy is limited by the noise
  level, it may be difficult to ascertain whether one should invoke
  complete latency or a gradual increase of the aging rate.

  Finally, Drummond and Israelachvili \cite{Dru1} have compared, for
  squalane and a star-shaped lubricant (PAO), aging under zero and
a finite shear stress close below the dynamic level. Contrary to what
was observed on PMMA/glass MCI, they find that, for a given $t_{w}$,
the stiction peak is systematically higher when waiting under zero
shear stress. This leads the authors to associate rejuvenation upon
sliding, responsible for the force drop producing the stiction peak,
with shear-induced molecular alignment, whose relaxation is certainly faster at
zero waiting stress. We will come back to this in the discussion of
$\S$III.F.

\subsubsection{Sliding dynamics}

\begin{figure}[h]
    \includegraphics[width = 7cm]{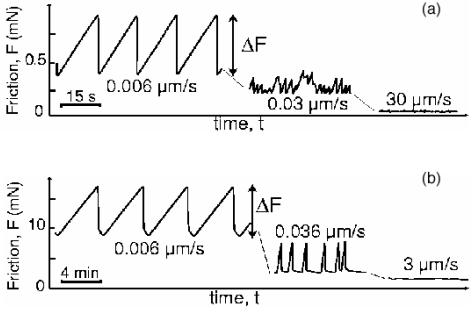}
    \caption{
    \label{fig:Gourdon}
    Dynamic friction force {\it vs} time for a squalane boundary
    lubrication layer driven at increasing velocities under pressures $P =
    1.4$ MPa (a); $5.4$ MPa (b) (adapted from Figure
    2 of ref. \protect\cite{Gour}).}
    \end{figure}

All BL systems with a finite static threshold exhibit the same
dynamical behavior in the explored low driving velocity domain (ranging
from $\lesssim 10^{-2} \mu$m/sec to a few $10 \mu$m/sec). Namely, as
illustrated by figure \ref{fig:Gourdon}~:

    --- For $V < V_{c1}$ periodic stick-slip (SS) motion.

    --- In a finite range $V_{c1} < V < V_{c2}$, the motion exhibits
    intermittency, in the form of randomly spaced stick-slip peaks
    \cite{Dru2}.

    --- For $V > V_{c2}$, these peaks disappear, sliding becomes
    steady \cite{Dru2}
    and, in this regime, dynamic friction is $V$-weakening.

In most cases studied, the amplitude of the SS force signal is
roughly constant up to $V_{c2}$~\footnote{ Whether or not such a discontinuous
transition exhibits hysteresis upon
increasing/decreasing $V$ is not documented,
to our knowledge.}.

It is worth mentioning that a qualitatively similar behavior is
observed with gelatin/glass extended sliding contacts \cite{gelatPRL}.
In this latter case it is associated with the fact that, in the SS
regime, sliding is spatially inhomogeneous -- it occurs via
propagating slip pulses (see $\S$III.C.4 below). In the
intermittent regime above $V_{c1}$, the corresponding large force
peaks are interspeded with smaller ones due to nucleation,
propagation and death of small slip events within the contact.
This remark suggests the possibility that sliding in BL junctions might
also exhibit spatial complexity. Such a model has been proposed by
Persson \cite{Bo}, who invokes the nucleation and propagation of
"shear-melted islands".

In an extensive study of squalane junctions, Gourdon and
Israelachvili \cite{Gour} have, in particular, investigated the effect
of normal load on the sliding dynamics. They find that the dynamics
described above prevails at high normal stresses ($P \gtrsim 5$ MPa).
However, at lower normal stresses ($¸ \lesssim 2$ MPa), a different
behavior is observed~: for $V \gtrsim V_{c1}$, the amplitude of the
erratic stick-slip decreases continuously down to zero.

Finally, using a SFA with a large sliding range ($500 \mu$m),
Drummond and Israelachvili \cite{Dru1} have studied, for squalane,
the shape of friction force transients following velocity jumps. These
transients are long-lived, and characterized by a distance comparable
with the contact diameter.

\subsection{Extended Soft Contacts~: Gelatin/Glass Friction}

They are particularly easy to realize with hydrogels, such as
gelatin, which exhibit both ultra-low shear moduli (typically a few
kPa) and an extended elastic regime. Their high solvent content
(typically $\gtrsim 80 \%$) endows them with specific frictional
properties resulting from their poroelastic character
\cite{DLJohnson}. For this reason, we separate this case from that of
non-swollen elastomers.\\

Gong and Osada \cite {Gong} have performed extensive studies of~:

    --- the influence on the dynamic friction level of the
    (attractive {\it vs} repulsive) nature of gel/substrate
    interactions. They have shown that repulsive interactions result
    in the formation of a thin interfacial layer of solvent, hence to
    hydrodynamic lubrication. For attractive couples, friction is
    strongly influenced by the formation of transient adhesive
    polymer-substrate bonds, which they modelized by adapting a
    description due to Schallamach (see below).

    --- the dependence of the friction force on normal loads in these
    different situations.

However, up to now, the only system on which the nature of the sliding
dynamics has been fully analyzed is gelatin on glass  \cite{gelEPJ},
on which we now concentrate.

\subsubsection{Static threshold}

The experiments described here were performed under zero normal load,
adhesion being strong enough to lead to solid friction, i.e. to a
finite static threshold.

  \begin{figure}[h]
    \includegraphics[width = 7cm]{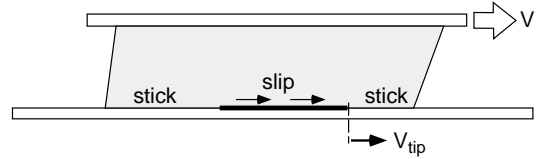}
    \caption{
    \label{fig:gel bloc}
    Schematic representation of a gel block driven at velocity $V >
    V_{c}$ (see text). The self-healing slip pulse propagates at
    velocity $V_{tip}$.}
    \end{figure}

When a gelatin block is sheared (Figure \ref{fig:gel bloc}), as
the force increases, the gel/glass contact first remains pinned, and the
block deforms elastically. At a threshold stress $\sigma_{0}$ sliding
sets in. This occurs via nucleation at the trailing block edge of an
interfacial fracture without measurable vertical opening, which
propagates forward with a constant velocity $V_{tip}$ in
the mm--cm/sec range, i.e. much smaller than that of transverse sound
(typically $\sim 1$ m/sec)
\footnote{Note that, while the shear modulus is
controlled by the elasticity of the loose polymer network, the
undrained bulk modulus is that of the solvent, so hydrogels can be
considered incompressible.}.
By changing the gelatin concentration, hence the mesh size $\xi$, and
the solvent viscosity $\eta_{s}$, it appears that $V_{tip} \sim
D/\xi$, with $D$ the collective diffusion coefficient \cite{DLJohnson},
measured from quasi-elastic light scattering experiments
\cite{Tanaka}~:
$D \sim G \xi^{2}/\eta_{s} \sim k_{B}T/\eta_{s}\xi$, with $G \sim
k_{B}T/\xi^{3}$ the shear modulus.

This indicates that the interfacial pinning, which must be severed in
order to crack, is provided by bonds with a typical lateral spacing
$d$ on the order of the mesh size.

Indeed, the value of the self-selected
$V_{tip}$ can be interpreted as follows. The energy released by bond
snapping events is emitted at the frequency $\omega \sim V_{tip}/d$,
and wave vector $q \sim d^{-1}$. The lowest deformation mode of the
gel block is the so-called Biot mode, associated with network
diffusion within the solvent $\omega_{D} = Dq^{2}$. It resonates with
emission, hence becomes most efficient to dissipate the fracture
energy, when $\omega_{D} \sim \omega$, from which it emerges that
  $d \sim \xi$
\footnote{It is worth mentioning that Rubinstein et al \cite{Fineberg}
heve recently observed that sliding between purely elastic solids also sets
in via propagation of interfacial cracks, but in the elastic case
these propagate at velocities in the sonic range. The low values of
$V_{tip}$ for gelatin are therefore, in contrast, a signature of
poroelasticity.}.

  \begin{figure}[h]
    \includegraphics[width = 7cm]{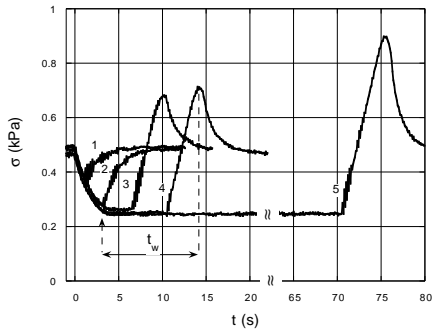}
    \caption{
    \label{fig:aging gel}
    Aging-induced growth of the static friction peak (curves $3,4,5$) for a $5
    \%$ gelatin gel/glass interface. If the drive is resumed
    before the interface resticks (left of vertical arrow)  no peak is observed
    (curves $1,2$).
    From ref. \protect\cite{gelEPJ}. }
    \end{figure}
	
Now, one also observes, as shown on figure \ref{fig:aging gel},
that the threshold $\sigma_{0}$ for crack
nucleation is not a constant, but increases logarithmically with
waiting time at rest $t_{w}$ (in the explored range $t_{w} \gtrsim 1 sec$).
This
indicates that adhesive bonding between the substrate and the
polymer tails and/or loops dangling from the disordered network
proceeds on two time scales~: a rather fast one, which leads to finite
static pinning with a bond areal density $\sim \xi^{2}$, followed by a
slow, thermally activated relaxation of the adsorbed polymer
configurations leading to the slow increase of this density.

\subsubsection{Sliding dynamics}

When the gel block is driven at velocity $V$, two types of dynamics
are observed.

\begin{figure}[h]
    \includegraphics[width = 7cm]{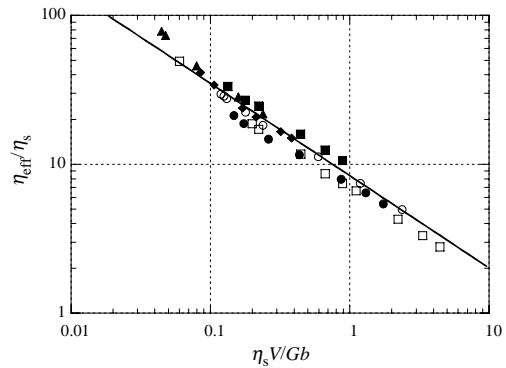}
    \caption{
    \label{fig:collapse}
      Reduced effective viscosity $\eta_{eff}/\eta_{s}$ versus Weissenberg
      number $We = \dot{\gamma}\tau_{R}$ for gelatin gels in
      water/glycerol solvents. Filled symbols : gelatin (wt \%)
     5(circles), 8 (squares), 10 
      (triangles), 15 (diamonds) in pure water. Open symbols: 5 wt \% 
      gelatin in 21 \% (circles) and 42 \%(squares) glycerol in water.}
    \end{figure}

    (i) Above a critical velocity $V_{c}$ (ranging, depending on gel
composition, from $\sim 25$ to $\sim 250 \mu$m/sec), sliding is
steady. The sliding stress $\sigma_{s}(V)$ is
velocity-strengthening, and corresponds to a shear-thinning rheology, namely~:

$$\sigma_{s}(V) \sim V^{1-\alpha}$$
with $\alpha = 0.6 \pm 0.07$. This is reminiscent of the behavior of
polymer solutions \cite{Bird}, and suggests that, in this sliding
regime, interfacial pinning plays only a minor role, dissipation
being due to the viscosity of a layer of thickness the mesh size (on
the order of $10$ nm)
formed by a solution of polymer segments attached to the network. If
such is the case, the shear strain rate $\dot\gamma
\sim V/\xi$, and the effective viscosity

$$\eta_{eff} = \frac{\sigma_{s}}{\dot\gamma} \sim \eta_{s} (\dot\gamma
\tau_{r})^{-\alpha}$$
with $\tau_{R}$ a typical relaxation time for a chain length $\sim
\xi$ in solution. Using for $\tau_{R}$ the Rouse time one gets, for a
variety of gelatin concentrations and solvent viscosities, the
collapse of data shown on Figure \ref{fig:collapse}, which lends
excellent support to the above interpretation.

    (ii) For $V < V_{c}$, periodic stick-slip prevails. Optical
observation shows that during the ``slip phase" sliding is not
homogeneous, but proceeds by propagation of {\it self-healing
pulses} (Figure \ref{fig:gel bloc}). Pulse heads are the previously
described crack tips, behind which the local slip velocity $v$
decreases from the standard quasi-diverging field towards the driving
value $V$. However, this regular decrease is suddenly interrupted
when $v$ reaches a value equal to the critical value $V_{c}$ which
turns out to coincide with the upper limit of the stick-slip regime.
At $v = V_{c}$, resticking (healing) occurs quasi-discontinuously.

When $V$ approaches $V_{c}$ from below, the pulse length increases, so
that the transition to steady sliding takes place via increasing time
spacing of pulses of quasi-constant amplitude, while being non
hysteretic. Moreover, for $V \simeq V_{c}$, an intermittent behavior,
already described above ($\S$III.C.3) is observed -- the interpulse
spacing becomes erratic, and they are interspeded with smaller force
signals associated to short-lived small propagating slip events. A
more detailed characterization of this complex dynamical behavior
will certainly be of interest.

\subsubsection{Rate and state interpretation}

Clearly, as for the previously considered cases, static strengthening
means interfacial aging at rest, which we expect to be associated with
rejuvenation, i.e. weakening, upon sliding, as proved by the
reproducibility of $\sigma_{0}(t_{w})$ in stop and go experiments.

This is consistent with the abruptness of the resticking process at
the trailing edge of the self-healing pulses. Indeed, assume that the
steady-sliding characteristics is $V$-weakening for $V < V_{min}$.
Close behind the pulse head, the local $v$, larger than $V_{min}$,
lies on the locally stable branch of $\sigma_{s}(V)$. As the distance
behind the head increases, $v$ decreases. When it reaches $V_{min}$,
the interface becomes unstable on a small spatial scale (smaller than
the optical resolution). In other words, since the driving stiffness,
provided by the very compliant gel block, is extremely low, the
interface jumps down to $V = 0$ at quasi-constant stress. This leads to
concluding that  the resticking velocity $V_{c}$ corresponds to a
minimum of $\sigma_{s}(V)$ above which, as appeared in the previous
paragraph, interfacial pinning should be negligible. This
interpretation is also consistent with the coincidence of the
resticking velocity and the disappearance of stick-slip as soon as
the asymptotic value $V$ of the local velocity $v$, needed for steady
sliding to get established, reaches the stable branch of $\sigma_{s}(V)$.

So, here again, it seems natural to try and define a structural age of
the gel/glass interface. Quite clearly, the corresponding dynamic
state variable must somehow measure the pinning energy associated
with the formation of polymer-glass adhesive bonds.

Already $50$ years ago, Schallamach \cite{Schall} proposed, in the context
of dry
rubber friction, a seminal model of such a dynamic state variable,
which remains the basis of all more recent theories of gel \cite{Gong}
and rubber \cite{Charitat} \cite{Chernyak} friction. In this model, briefly
summarized an discussed in Appendix D, the elementary mechanical instability
(see
$\S$II.C.4) is the snapping of a bonded molecule out of its adsorption
site under shear loading through an elastic spring (here the polymer
segment). Once this energy barrier has been jumped over, under the
combined effect of loading and thermal activation, the molecule is
freely advected until it readsorbs via a thermally activated process.
Of course, readsorption is thwarted by advection, which limits the
time available for this process to take place. The state variable is
thus the density of bonded molecules. The friction stress is the
product of this $V$-weakening density, and of the $V$-strengthening
average depinning force (the faster advection, the less time for
escaping above high barriers). So it exhibits a bell shape with a maximum
at $V = V_{max}$. The low velocity regime is an Eyring
viscous one, with the bond density close to its equilibrium
saturation level. In the $V$-weakening high velocity regime the
depinning force saturates, $\sigma_{s}(V)$ is then fully controlled
by dynamic rejuvenation, hence vanishes for $V \rightarrow \infty$.

Models of the Schallamach type call for two important remarks~:

    {\it (i)} The extent of the Eyring-like low $V$ regime ($V <
\tilde{V}$, see Figure \ref{fig:tout Schall})
crucially depends on the height of the desorption
barrier. When, as often assumed -- explicitly or not -- it is on
the order of a few $k_{B}T$, $\tilde{V}$ is non-negligible. A strong
consequence is that, in such cases, no static threshold should be
observable except if one could load at exceedingly large velocities.
This points towards the interest of systematically associating
studies of dynamic friction in such systems with stop and go
experiments.

Gelatin/glass interfaces exhibit well defined static force peaks and
crack tips. We must therefore conclude that, in this case, we are
dealing with strong bonds -- in which case $\tilde{V}$ becomes negligibly
small (see end of $\S$II.C.4).

    {\it (ii)} If we stick to the Schallamach prediction, in the case
of gelatin, $\sigma_{s}(V)$ should be uniformly $V$-weakening, since
in the explored range $V \gg V_{max}$. How can one then explain the
existence of the $V > V_{c}$ strengthening behavior?

We meet here with an important shortcoming of the Schallamach model.
Namely, unpinned molecules are assumed to glide freely, that is,
viscous dissipation in the interfacial layer is omitted - although,
since its thickness is typically nanometric, shear rate levels are
high. In the case of gelatin we saw that, on the contrary, it is this
contribution which fully accounts for the observed non-newtonian
friction at the essentially depinned interface.
This illustrates the importance of taking into account in this kind of problem,
not only interfacial pinning, but, as well, junction viscosity. When
both effects are of comparable importance, this open problem becomes a
true nanofluidics one, as far as the size of the confined polymer chains is
precisely comparable with the thickness of the ``channel" , i.e. the
junction.

In summary, once complemented with standard viscous dissipation in the
sliding junction, the Schallamach picture, which identifies the state
variable with the density of adsorbed bonds, provides a sound basis on
which to modelize gel friction. Clearly, a more elaborate modelization
of the dangling chains, taking into account the existence of a
multiplicity of possible bonding configurations \cite{Semenov} and a
detailed description of chain dynamics \cite{Charitat} would be needed
to make such models more quantitative, as well as to touch upon the
question of long term aging.

\subsection{Extended Soft Contacts~: Dry Elastomer Friction}

Due to its importance in a number of applied fields (e.g., to name
only a few, traction of tires, windshield wiper efficiency and
durability), this question has been a subject of very active
investigation in the past six decades. We restrict ourselves here to a
sketchy summary of the commonly accepted concepts used for
interpretation of interface geometry {\it vs} molecular adhesion
effects. For extensive bibliographies, see \cite {Roberts} \cite{Savkoor}
\cite{Chaudhury}.

We consider here the case of elastomers in extended contact with much
stiffer solids, which can thus be assumed non deformable. Bulk
viscoelastic dissipation is very important in rubbers. Its strong
temperature dependence follows the so-called WLF time-temperature
superposition principle \cite{Ferry}. Grosch \cite{Grosch}, and
Ludema and Tabor \cite{Ludema}, have shown that one must distinguish
between contributions to the sliding friction stress governed
respectively by bulk and interfacial dissipation.

\subsubsection{Bulk dissipation}

It comes into play as soon as the stiff partner surface exhibits non
negligible roughness - on a scale larger than the average cross-link
spacing. Assume for a moment that the roughness has a characteristic
wavelength $\lambda$. Then, in stationary sliding, a given material
point within the elastomer feels a stress field modulated at a
frequency $\omega \sim V/\lambda$, which penetrates down to a depth
$\sim \lambda$ below the surface. This results in a dissipation
governed by the loss modulus of the bulk, $G''(\omega)$, i.e. obeying
a $V-T$ superposition principle. One is thus able to account for the
most salient feature of rubber friction, namely the existence of a
maximum of the $\sigma_{s}(V)$ curve, at a velocity $V_{m}$, which
increases with the temperature $T$, above which stick-slip motion
sets in \cite{Grosch}.

\subsubsection{Interfacial dissipation}

Grosch \cite{Grosch} identified, in his experimental results on
friction of several elastomers on silicon carbide paper, an additional
feature -- a shoulder on the $\sigma_{s}(V)$ curve at a velocity
$V_{m} \simeq \omega_{m}\Lambda$, where $\omega_{m}$ corresponds to
the maximum of the loss modulus and $\Lambda \sim 6$nm. This shoulder
turned into a broad maximum when sliding on smooth ``wavy" glass. He
attributed this to interfacial dissipation -- i.e. to the snapping
out of advected adhesive bonds as modelized by Schallamach (see
$\S$III.D.3).

It is only recently that progress in surface control enabled
Vorvolakos and Chaudhury (VC) \cite{Chaudhury} to reconsider this
question. They used two model systems, constituted of cross-linked
PDMS sliding on {\it (i)} a self-assembled silane monolayer supported
by a Si wafer {\it (ii)} a thin film of glassy PS. They were thus able
to get rid of surface roughness effects, hence of bulk dissipation.

\begin{figure}[h]
    \includegraphics[width = 7cm]{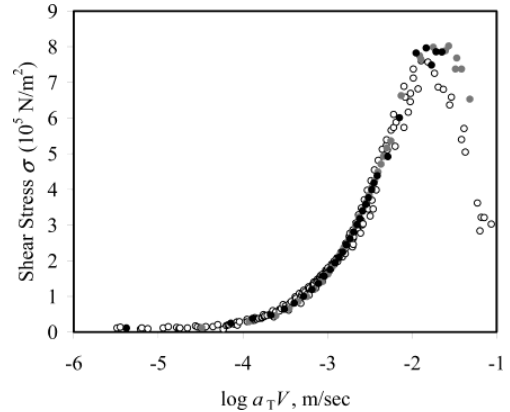}
    \caption{
    \label{fig:WLF}
    Temperature-dependent friction stress data for PDMS on a silanized
    silicon surface shifted to room temperature using a WLF shift factor.
    Open, gray and black circles represent data at $298, 318$ and $348$ K
    respectively. From ref. \protect\cite{Chaudhury}.}
    \end{figure}

Silane coverage results in very low energy surfaces, so one expects
polymer substrate adhesive bonds to be rather weak in this case and,
hence, Schallamach's model to be appropriate. Indeed, VC show that
their results for $\sigma_{s}(V)$ in this system (see figure
\ref{fig:WLF}), which extend over
$5$ velocity decades, are well explained qualitatively in this frame
\footnote{ Interfacial viscous dissipation is neglected in the VC
analysis. As already mentioned, it should come into play, especially
in the $V$-weakening regime $V > V_{m}$, where it is certainly needed
to interpret the observed stick-slip behavior.}. Moerover, they deduce
from a time-temperature superposition argument an energy scale $\sim
25$ kJ.mol$^{-1}$, i.e. $\sim 0.3$ eV/bond. Whether this energy, much
larger than the van der Waals ones expected for PDMS/silane
interactions, is relevant to adhesive bond snapping
in this system remains an open question.

PDMS bonding to PS is likely to be stronger than that on silane.
Indeed, consideration of the VC results in the low velocity limit
leads to raising the question of the possible existence of a finite
static friction threshold. Systematic investigation of the transient
behavior at the onset of sliding would certainly be enlightening.\\

Finally, it is worth mentioning that non homogeneous sliding has been
observed \cite{Schallwaves} \cite{Roberts}  at some smooth soft
rubber/glass interfaces. This occurs via propagating self healing
pulses. But, contrary to the gelatin case where contact is maintained
in the sliding regions, these {\it Schallamach waves} are of the mode-I
fracture type. That is, they consist of regions of interfacial
detachment with macroscopic opening which re-adhere at the back edge.
So, in this regime, dissipation is essentially controlled by adhesion
hysteresis \cite{Barquins}. Though high compliance
and large viscoelastic losses are thought to promote this sliding mode,
no prediction relating material properties to the dynamics, nor even
to the occurrence of Schallamach waves is available up to now.

\subsection{A tentative classification~: Jammed junction plasticity
vs adsorption controlled dynamics}

The various frictional behaviors briefly described above pertain to
junctions which differ in several respects, namely~:

$\bullet$ Normal stress level, i.e. confinement pressure $P$, which
ranges from zero (gelatin) to about the yield stress of the softer
bulk material (rough on flat MCI).

$\bullet$ Level of adhesive interactions, i.e. strength of the pinning
sites provided by the confining potential. This ranges from very weak
(fully silanized substrates) to strong (e.g. bare glass).

$\bullet$ Density of adsorbable units, ranging from dense (polymer
glasses) to (semi)dilute (hydrogels).

All these systems share two properties~:

    -- they exhibit a static threshold~\footnote{ Except possibly for
    some elastomers under low normal stress and in contact with highly
    passivated substrates.}

    -- this threshold ages slowly with time spent at rest.

The experimental signature of this structural aging is the slow (in
general quasi-logarithmic) growth of the upper limit of the elastic
regime. In experiments where the slider is remotely driven at constant
velocity $V$, it results in general in a force peak. This unambiguously
reveals rejuvenation upon sliding, since the sliding velocity $v$ is
equal to the driving one $V$ both in the steady sliding state and
at the peak, where the slid distance is minute, so that the age is
basically given by the waiting time before pulling~\footnote{ When
established sliding is not steady, but occurs via stick-slip, static aging
results in the growth of the {\it first} peak with stopping time.}.
This means that the state of the sliding junction is specified, not
only by its instantaneous velocity, i.e. by the shear rate
$\dot{\gamma}$ it experiences, but also by the value of some
dynamical state variable.

  On the other hand, different classes of systems exhibit different
  steady state rheologies, from logarithmic strengthening for highly
  confined PMMA/glass to power law for gelatin.

  While these behaviors call for rate and state descriptions, it is
  clear that various classes of systems must be distinguished, in
  terms of the microscopic nature of the state variable $\varphi$,
  which is necessarily related with the junction structure and,
  hence, with the pinning mechanism giving rise to static friction. The
  existing body of experimental results discussed in $\S$III.B--E is
  still far from sufficient to provide a firm basis for such a
  phenomenology. So, the classification which we propose here is only
  tentative, and primarily intended to help structuring further questions to
  be, in a first stage, investigated experimentally.\\

  Let us start from the very high confinement regime, realized in the
  MCI configuration ($\S$III.B) where, for PMMA, the confining
  pressure $P \sim 100$ MPa is comparable with the bulk yield stress.
  We saw that junctions between such a polymer glass and highly
  silanized glass behave as soft glassy nanometer-thick media.

  In other words, we claim that, in such a case, the disordered
  material of the dense junction jams \cite{livre Nagel}, i.e.
  solidifies under
  confinement. It ages as a glass, by relaxing towards deeper local
  minima of the energy landscape (inherent states) and, when it flows,
  dissipation can
  be attributed to irreversible flips of molecular-sized shear
  transformation zones (STZ) \cite{Falk}. In this case, {\it frictional
  sliding is nothing but plastic flow of this interfacial glass}
  which, being weaker than the bulk, naturally localizes shear.

\begin{figure}[h]
    \includegraphics[width = 7cm]{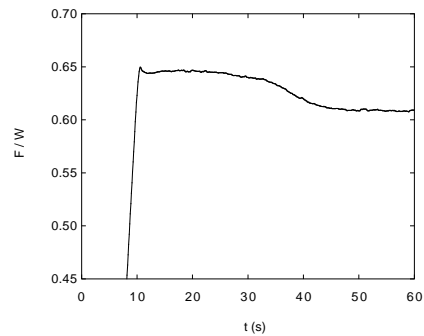}
    \caption{
    \label{fig:trans long}
    Reduced friction force versus time after a $300$s stop under zero
    shear stress. The rough PMMA slider is driven at $V = 10 \mu$m/s on a
    poorly silanized track.}
    \end{figure}

  Now, note that, in the MCI experiments, when the glass substrate is
  well silanized, no direct manifestation of the polymeric nature of
  the junctions is observed. However, when the same experiments are
  performed on a poorly silanized substrate \cite{Lionel EPJB}, the
  previously observed rather narrow aging peak persists, but is followed
  by a long bumpy stress transient (see Figure \ref{fig:trans long}).
  Moreover if, after reaching the steady state, one performs
  stop-and-go tests with a high stopping shear stress level, the long
  transient is absent when restarting motion. If, on the contrary, the
  stop is performed under zero shear stress,  the wide bump
  reappears. This strongly suggests that these transients, associated
  with the presence of stronger corrugations of the confining substrate
  potential, are due to gradual stretching and alignment of the
  constitutive molecules, which persists when the shear stress is
  maintained, and relaxes otherwise. Such rubbing-induced alignment
  has been documented already long ago by Pooley and Tabor
  \cite{Pooley-Tabor} on teflon, a material for which it is
  particularly strong. It would be desirable, following these authors,
  to test this
  interpretation by devising a set-up which would permit to change the
  sliding direction by e.g. $90 ^{\circ}$.\\

  This observation naturally raises the question of the {\it competition
  between jamming and adhesive pinning to the substrate}. It is reasonable to
  expect that, the lower the confining pressure $P$, the weaker the
  jamming, i.e. the relative contribution of STZ flips to dissipation.

  In boundary lubrication experiments performed in the SFA, pressure
  levels are smaller that those in MCI junctions by at least two orders
  of magnitude. As mentioned in $\S$III.C.2, the amplitude of the
  transients observed on squalane by Drummond and Israelachvili
  \cite{Dru2} is
  larger when stopping under smaller shear stress. They also seem to
  be associated with quite broad stress bumps. This leads one,
  following the authors, to associate these transients with molecular
  alignment. Whether or not a narrow glassy-like aging peak is
  observable at the highest SFA pressures would be worth investigating
  in detail.

  This remark points toward the need for trying to bridge, for given
  interfacial materials, between SFA and MCI pressure levels, for
  example with the help of the ball-on-flat configuration. Such
  comparisons are necessary in order to appreciate to which extent SFA
  data can be safely extrapolated to the conditions prevailing in
  macroscopic contacts between hard solids.\\

  The other limiting situation of purely adhesive pinning. It
  is exemplified by gelatin gels in
  contact with glass under zero or very small \cite{gelEPJ} normal
  load ($\S$III.D). In this case the junction is constituted of a (semi) dilute
  solution of protein segments in water -- no jamming can be invoked.
  The pinning responsible for the finite static threshold is due to
  adsorption of the polymer molecules onto the glass substrate, and
  one may, at least semi-quantitatively, identify the state variable
  $\varphi$ with the areal density of adsorption bonds. Such a structure
  is able to relax by a combination of thermal desorption,
  chain configuration rearrangements and readsorption which leads, at
  rest, to the slow increase of $\varphi$. The number of candidate
  sites on a segment of course depends on the chemical nature of the
  network-forming polymer chains. This dynamical problem, although non
  trivial, might be amenable to a theoretical treatment which would
extend equilibrium studies such as those of Subbotin et al. \cite{Semenov}
  along the lines of the Schallamach-like steady state model built by
Charitat and
  Joanny \cite{Charitat} for the case where only the end monomer is
  adsorbable.

  In such a completely non-jammed junction, frictional dissipation
  contains two contributions due respectively to~:

     --- Thermally helped advection-induced depinning.  It is this
     mechanism which controls the solid friction behavior proper for
     such {\it adsorption-controlled interfaces}.

     --- Viscous flow of the polymer-water mixture in the sheared
     interfacial layer.

While the latter is velocity-strengthening, the former decreases with
decreasing $\varphi$, so it is basically velocity-weakening.\\

Elastomers probably correspond to an intermediate situation. The
interfacial junctions which they form certainly have a dense polymer
content, while clearly not being solid (jammed), since their bulk
itself is not. Junction thickness scales as the distance between
entanglements. One of the open questions is that of evaluating
realistically the viscous contribution to friction of such sheared
layers. Whether they exhibit, as might be expected, slow aging at rest
is not documented yet, as far as all experiments on elastomer friction
have focussed up to now on the steady sliding behavior and on its
temperature dependence.

\section {Conclusion}

The route we have chosen to follow in this review has led us from
macroscopic dynamics down to gradually decreasing space scales. Solid
friction thus appears as a particularly favorable area of material
science in which the governing mechanisms are ``universal" enough to
allow for a rather general phenomenology down to the nanoscopic
level.\\

A first level of analysis, relevant to the widespread case of
macroscopic interfaces between rough hard solids, identifies two main
physical ingredients~:

   (i) A shear-induced depinning process, in which mechanical
   instabilities lead to structural rearrangements of nanometric
   shear transformation zones within the junctions where shear
   localizes. It is the associated multistability of these glassy
   junctions which is responsible for the existence of finite
   friction thresholds.

   (ii) Geometric aging, i.e. slow creep growth of the area of the
   sparse microcontacts forming such interfaces under the high contact
   pressures associated with this geometry. Sliding limits the
   duration of contact life, making geometric age a dynamical
   variable with a memory of the sliding history.

This description provides the physical basis of the Rice-Ruina
constitutive laws which, when properly extended, account for all the
main features of the low velocity frictional dynamics. It also permits
to point more precisely the limits of such a constitutive law. Among
these, an important one is concerned with higher velocities ---
typically $\gtrsim$ a few $100 \mu$m/sec -- at which geometric age
becomes small, leading to the saturation of the destabilizing
$V$-weakening behvior of dynamic friction, and to its inversion into
$V$-strengthening.
Behaviors in the intermediate (mm/sec to cm/sec) range are still
insufficiently documented experimentally to allow for detailed models
of dissipation in this regime.

For still faster sliding, self-heating becomes non negligible, turning
temperature into a new relevant state variable which affects junction
rheology. An example of such an effect is that of the stick-slip
dynamics of the bowed violin string, which has been modelized by Smith
and Woodhouse \cite{Woodhouse} in terms of the triggering by slip-induced
heating  the
of a transition from sticking solid to
viscous fluid of the rosin rubbed on the bow. \\

Beyond the slow dynamics leading to the notion of geometric age,
another one emerges, associated with structural aging of the
junctions, whose effects are masked, for rough-on-rough systems, by
the larger geometric aging ones. Structural aging at rest and its
counterpart -- rejuvenation when sliding -- is at work in all types
of frictional junctions, leading to slow growth of static thresholds
with time and to stick-slip-like dynamical instabilities. It affects
the topography of the multistable energy landscapes which govern frictional
rheology. Further experimental information will be needed in order to
try and specify it in terms of precise state variables, whose nature
is probably not unique.

On the basis of comparisons between different classes of interfaces,
we have suggested a schematic classification in terms of two limiting
types.

   (i) {\it Jammed junctions}, solidified under confinement into a soft
   glass structure. They are relevant at high contact pressures such
   as met with multicontact interfaces. Multistability results from
   their glassy structure itself, and in this case frictional
   dissipation enters the wider class of problems constituted of
   plasticity of amorphous media and soft glass rheology.

   (ii) {\it Purely adhesive junctions} for which pinning is due
   to adsorption bonds between the junction molecules and the
   substrates(s). They form at contacts under low confinement pressures
   involving gels and, more generally, polymers.

Many real junctions probably are a compound of these two ideal types,
 their frictional dissipation containing contributions of both
types whose respective weights depend on pressure, temperature,
density and on the physico-chemical nature of the confining
surfaces.\\

While, in our opinion, the question of geometric age effects in
multicontact solid friction is now reasonably cleared up, that of
juction rheology remains a largely open issue worth of further
investigation. In order for modelization to make significant progress,
more detailed experimental information is needed, in particular about~:

   -- Systematic trends associated with varying pressure and
temperature, for systems with controlled surfaces.

   -- The effect of various aging histories, paralleling analogous studies
in structural and spin glasses. In the case of adsorption-controlled
junctions, information about aging might also be obtained via
adhesion tests \cite{adhesion}, provided that these be performed on
materials with negligible viscoelasticity.

Numerical simulations of sheared junctions in the presence of
realistic confining potentials should also be of great help.\\

Finally, the reader may have been surprised that we never used terms
such as ``nanofriction" or ``nanotribology". This we did purposefully,
since we consider them rather misleading. Indeed, ``nanotribology" is
often associated with investigations of the nature of the basic
dissipative processes which, as we saw, occur on the nanometric
scale. However, friction, as defined in terms of a stress or a
friction coefficient, results from the self averaging of such events
within contacts of, at least, mesoscopic (micrometric) size. This is
precisely why friction laws do not make sense on the nanometric scale.

We consider that the term nanofriction should be reserved for
situations where lateral contact sizes are themselves of nanometric
order, in which case different effects may emerge, such as
exemplified by aging due to capillary condensation \cite{Riedo}. It
is also likely that, in such regimes, the nature of dissipative
processes \cite{Kim} is different from that in mesoscopic contacts.

\section*{Acknowledgments}

We are particularly grateful to J.R. Rice, K.L. Johnson and Y. Br\'echet
for encouraging our first steps in the field of friction, and letting us
benefit liberally from their vast knowledge of mechanics and material
science.
We are deeply indebted to P. Nozi\`eres, B. Perrin, B. Velicky and F. Heslot
for their precious contributions at various stages of our work on this
subject, as well as to
L. Bureau and O. Ronsin for a long lasting collaboration, and fruitful
exchanges on these and related subjects.


\section* {Appendix A~: Shear stiffness of a MCI}

Already long ago, a number of studies have tried to get independent
information about the structure of MCI from the load dependence of
various interfacial properties besides friction. These include
electrical \cite {Tabor} and thermal \cite{thermal} conductances,
variations of closure with normal
load \cite{closure}, acoustic reflection and interfacial normal and transverse
mechanical stiffnesses \cite {ultrasound}. In order to confront experimental
results with the
Greenwood-Williamson model, each of these properties must be
calculated within this same frame.

Let us sketch out, as an example, the evaluation of the interfacial
shear stiffness. For a single Hertz contact of radius $a$ between
two identical spheres of shear modulus $G$ and Poisson ratio $\nu$, it
was calculated by Mindlin \cite{Mindlin} \cite{Johnson} to be~:

\begin{equation}
\label{eq:A1}
k_{s} = \frac{4Ga}{2 - \nu}
\end{equation}
which expresses the fact that the strain energy due to shearing is
essentially localized, in each medium, within a volume of section
$\sim a^{2}$ and
height $\sim a$. So, the relative shear displacement involved in the
definition of $k_{s}$ is that between any two points on both sides of
the contact at distances $\gg a$ from it.

Consider a set of $n$ GW microcontacts between two blocks of
thickness $D$, shear modulus $G$. One may separate this elastic system
into an interfacial layer of thickness $\gg a$ but $\ll D$ sandwiched between
two bulk regions of thickness $\simeq D$. As far as $\Sigma_{R} \ll
\Sigma_{app}$, the microcontacts are elastically independent, and the
total {\it interfacial stiffness} $K_{s}$ is simply the sum of the
individual ones, i.e., with the notations of $\S$ II.C~:

\begin{equation}
\label{eq:A2}
K_{s} = \frac{4G}{2 - \nu} n \bar{a}
\end{equation}
From equations (\ref{eq:GW4}), (\ref{eq:GW5})

\begin{equation}
\label{eq:A3}
K_{s} = \frac{4G}{(2 - \nu) \sqrt{\pi} E} \frac{W}{s} = \frac{2}{(2 -
\nu)(1 + \nu)} \frac{W}{s}
\end{equation}
So, the GW model predicts, contrary to immediate intuition, that the
interfacial stiffness of an elastic MCI is independent of the bulk
material shear modulus.

Of course, experiments measure the compound response of the interface
and the bulk. A more important warning comes from the fact
that expression (\ref{eq:A1}) is valid only in the limit of very small
ratios of shear to normal forces. Indeed, Mindlin has shown \cite
{Johnson} that increasing
shear induces, at the periphery of the contact, a slipping annulus
along which, would it remain non-slipping, the friction sliding
threshold would be overcome. This effect results in a non-linear
weakening behavior of $k_{s}$ \cite{Johnson} and of the MCI
response  \cite{Lionel Proc}.

Consider now a GW interface in the fully plastic limit. When a small
shear force is applied for the first time, it induces some further
plastic flow, probably leading to some small adaptation of contact
radii. Under subsequent unloading-reloading cycles, these adapted
contacts respond elastically, with an individual stiffness $\sim
Ga$. $K_{s}$ is then immediately evaluated to be  $\sim W/\lambda$,
where the length $\lambda$ is of the order of $s/\psi$, with $\psi$
the plasticity index defined in section II.C \cite {Pauline Proc Roy Soc}.

The experiments of Berthoud and Baumberger \cite{Pauline Proc Roy Soc} on
PMMA (a polymer glass)/PMMA and AU4G (an aluminum alloy)/AU4G have
confirmed the Amontons-like behavior (\ref{eq:A3}) of $K_{s}$. The
lengths $\lambda$ for the two systems were different, but both on the
order of a few microns. They were found to be compatible, in order of
magnitude, with the values predicted from the above model.
These results confirm the robustness of the three properties of MCI,
listed in section II.C, which emerge from the GW model.

In view of the difficulty of determining accurately $s$ and, even
more, $R$, as well as of the numerous approximations involved in the
GW model and, even more, of its elasto-plastic extension, it would be
fallacious to expect a more quantitative agreement.

Amontons-like behaviors have also been measured to hold, with various
degrees of accuracy depending on the specific difficulties inherent to
each experimental method, on the other interfacial properties listed at
the beginning of this Appendix.

\section* {Appendix B~: The viscoelastic GW model}

Our derivation summarizes the work of Hui et al \cite{Hui}.

Let us consider a MCI between a linear viscoelastic, incompressible
and isotropic medium and a rigid one -- a good approximation for
contact between a rough elastomer and a rough hard solid. We want to
describe this MCI in the the frame of the simplest version of the GW
model (exponential distribution of summit heights).

In a linear viscoelastic material, shear stress and strain are
related via a retarded elastic shear modulus $G(t)$, by the
constitutive equation~:

\begin{equation}
\label{eq:B1}
\sigma(t) = \int_{- \infty}^{t} dt'\,G(t - t')\,\frac{d\epsilon
(t')}{dt'}
\end{equation}
which can be formally inverted into

\begin{equation}
\label{eq:B2}
\epsilon (t) = \int_{- \infty}^{t} dt'\,J(t - t')\,\frac{d\sigma
(t')}{dt'}
\end{equation}

$J(t)$, which characterizes the strain response to a unit stress step
applied at $t = 0$, increases with time, as a result of the progressive
relaxation of the constitutive polymer molecules.

Let us first consider the viscoelastic analogue of the single Hertz
contact ($\S$ II.A.2), created by applying the load $w$ at time $t = 0$.
It is intuitively clear that, since the compliance increases, the
compression $\delta(t)$ and the contact radius $a(t)$ increase monotonously.
Lee and Radok \cite{Lee-Radok} have shown (see also \cite{Johnson})
that in such a case the solution of the contact problem is that
deduced from the Hertz one by simply replacing the shear modulus by
its retarded counterpart $G(t)$. So, the Hertz load-radius relation
(eq. (\ref{eq:Hertz})) becomes~:

\begin{equation}
\label{eq:B3}
w = \int_{- \infty}^{t}\, G(t - t') \frac{d}{dt'}\left[\frac{4}{\pi
R^{*}}\,a^{3}(t')\right]
\end{equation}
with $a^{2}(t') = R^{*} \delta(t')$.

We now turn to the corresponding GW problem, contact under the normal
load $W$ being first established at $t = 0$. As time increases,
clearly, the interfacial separation $d(t)$ will decrease
monotonously, the number of microcontacts $n(t)$ and their individual
area increase, leading to the increase of the real area of contact
$\Sigma_{r}(t)$.

The three GW equations (eqs.(\ref{eq:GW1}) - (\ref{eq:GW3})) become~:

\begin{equation}
\label{eq:B4}
n(t) = N \int_{d(t)}^{\infty}  dz \frac{1}{s}\,e^{-z/s}
\end{equation}

\begin{equation}
\label{eq:B5}
\Sigma_{r}(t) = N \int_{d(t)}^{\infty} dz \pi R(z-d(t)) \frac{1}{s}\,e^{-z/s}
\end{equation}

\begin{widetext}
\begin{equation}
\label{eq:B6}
W(t) = N \int_{d(t)}^{\infty} dz \frac{1}{s}\,e^{-z/s}
\int_{-\infty}^{t} dt'\,\theta(t' - t_{0}(z))
G(t-t')\,\frac{d}{dt'}\left(\frac{4}{3}R^{1/2}\left(z-d(t')\right)^{3/2}\right)
\end{equation}
\end{widetext}

where $\theta$ is the unit step function, $W(t) = W\theta(t)$, and the
function $t_{0}(z)$ is the inverse of $d(t)$, i.e.

\begin{equation}
\label{eq:B7}
d\left(t_{0}(z)\right) = z
\end{equation}

Equation (\ref{eq:B6}) is based upon relation (\ref{eq:B3}) and
states that each contact with summit height $z$ existing at time $t$
has contributed since
$t' = t_{0}(z)$ when it first appeared, i.e. for $t' >
t_{0}(z)$ or equivalently, $ d(t') < z$.

Setting $y = \left(z - d(t')\right)/s$ and performing the
$y$-integration in eq.(\ref{eq:B6}) yields~:

\begin{equation}
\label{eq:B8}
W(t) = N\left(\pi R s^{3}\right)^{1/2} \int_{-\infty}^{t}
dt'\,G(t-t')\frac{d}{dt'} \left(e^{-d(t')/s}\right)
\end{equation}
which can be inverted, with the help of definitions (\ref{eq:B1}),
(\ref{eq:B2}) into~:

\begin{equation}
\label{eq:B9}
e^{-d(t)/s} = \frac{1}{\left(\pi R s^{3}\right)^{1/2}} \int_{-\infty}^{t}
dt'\,J(t-t') \frac{dW(t')}{dt'}
\end{equation}

On the other hand, from equation (\ref{eq:B5})

\begin{equation}
\label{eq:B10}
\Sigma_{r}(t) = N\pi R e^{-d(t)/s}
\end{equation}
so that, finally, since $dW/dt' = \delta(t')$, the real area of
contact evolves according to~:

\begin{equation}
\label{eq:B11}
\Sigma_{r}(t) = \sqrt{\frac{\pi R}{s}}J(t) W
\end{equation}

\section*{Appendix C~: The driven block~: Linear stability analysis}

Consider the system defined in $\S$II.D.2, made of a block of mass
$M$, carrying the normal load $W$, driven at the constant velocity $V$
through a spring of stiffness $K$
\footnote{We assume that, as is commonly the case, $K$ is much
smaller than the interfacial stiffness. Finite interfacial compliance
effects are evaluated in \cite {Cochard}.}.
The equations describing the motion read~:

\begin{equation}
\label{eq:C1}
M \ddot{x} = - K\left(x - Vt - \alpha\right) - W \mu_{d}(\phi, \dot{x})
\end{equation}

\begin{equation}
\label{eq:C2}
\dot{\phi} = 1 - \frac{\dot{x}\phi}{D_{0}}
\end{equation}

In the RR model~:

\begin{equation}
\label{eq:C3}
\mu_{d} = \mu_{d0} + B \ln\frac{\phi}{\phi_{0}} + A \ln\frac{\dot{x}}{V_{0}}
\end{equation}
with $V_{0}$ some reference velocity, $\phi_{0} = D_{0}/V_{0}$.
However, in view of the shortcomings of expression (\ref{eq:C3})
discussed in $\S$II.D, we will leave the functional form of $\mu_{d}$
unspecified at this stage.

Equations (\ref{eq:C1}), (\ref{eq:C2}) have the stationary solution~:

\begin{equation}
\label{eq:C4}
x_{st}(t) = Vt - \alpha - \frac{W}{K} \mu_{d}\left(\frac{D_{0}}{V},
V\right)
\end{equation}

\begin{equation}
\label{eq:C5}
\phi_{st} = \frac{D_{0}}{V}
\end{equation}

In order to study its (linear) stability, we linearize equations
(\ref{eq:C1}), (\ref{eq:C2}) in $\delta x(t) = x(t) - x_{st}(t);
\delta\phi(t) = \phi(t) - \phi_{0}$. Then~:

\begin{equation}
\label{eq:C6}
\frac{M}{W}\delta \ddot{x} + \frac{K}{W}\delta x +
\frac{\mu_{\dot{x}}}{V} \delta\dot{x} + \frac{V\mu_{\phi}}{D_{0}} \delta
\phi = 0
\end{equation}

\begin{equation}
\label{eq:C7}
\delta\dot{\phi} + \frac{V}{D_{0}} \delta\phi + \frac{1}{V}
\delta\dot{x} = 0
\end{equation}
with$\mu_{\phi}
= \partial\mu_{d}/\partial(\ln\phi)$,
$\mu_{\dot{x}} = \partial\mu_{d}/\partial(\ln\dot{x})
V$, both quantities being evaluated at $\phi = D_{0}/V,\,\dot{x} = V$.
Note that we saw that, for slowly sliding MCI, $\mu_{\dot{x}} > 0$,
$\mu_{\phi} < 0$, and $\mu_{\phi} - \mu_{\dot{x}} > 0$. The solutions
of this system are of the form~:

\begin{equation}
\label{eq:C8}
{\delta x \choose \delta \phi} = {\xi \choose \psi} e^{i\Omega t}
\end{equation}
and the frequencies $\Omega$ of these eigenmodes are the roots of~:

\begin{widetext}
\begin{equation}
\label{eq:C9}
-i\frac{M}{W} \Omega^{3} - \left(\frac{\mu_{\dot{x}}}{V} +
\frac{MV}{WD_{0}} \right) \Omega^{2} + \left( \frac{K}{W} +
\frac{\mu_{\dot{x}} - \mu_{\phi}}{D_{0}}\right) i\Omega + \frac{K}{W}
\frac{V}{D_{0}} = 0
\end{equation}
\end{widetext}

If, for all roots, $Im \Omega > 0$, all (infinitesimal) fluctuations
about the stationary state decay, steady sliding is (locally) stable.
For very large $K$ the solutions of (\ref{eq:C9})
read~:

\begin{equation}
\label{eq:C10}
\Omega_{1} \simeq \frac{iV}{D_{0}}\,\,\,\,\,\,\,\, \Omega_{\pm} =
\pm \sqrt{\frac{K}{M}} + (1 + 3i) \frac{W\mu_{\dot{x}}}{MV}
\end{equation}
Since $\mu_{\dot{x}} > 0$, the system is stable in the large $K$ limit.
As the stiffness decreases, it becomes unstable if, and when, the
imaginary part of one at least of the roots vanishes. One easily
checks on equation (\ref{eq:C9}) that this occurs when~:

\begin{equation}
\label{eq:C11}
\frac{K}{W} = \left(\frac{K}{W}\right)_{c} = \left(\mu_{\phi} -
\mu_{\dot{x}}\right)\left( 1 +
\frac{MV^{2}}{WD_{0}\mu_{\dot{x}}}\right)
\end{equation}

As expected, instability only occurs when $\mu_{\phi} -
\mu_{\dot{x}} > 0$, i.e. (see $\S$II) when the steady sliding
$\mu_{d}$ is velocity-weakening, which is indeed the case for our
systems in the $V$-range under consideration.

Note that inertia only comes into play in eq.(\ref{eq:C11}) via the
quantity $\theta = (MV^{2}/WD_{0}\mu_{\dot{x}})$. For ablock sliding
under its own weight ($W = Mg$), with $D_{0} \gtrsim 1 \mu$m,
$\mu_{\dot{x}} \approx 10^{-2}$, $\theta \lesssim 10^{-5}\left(V_{\mu
m/sec}\right)^{2}$ can safely be neglected in our low velocity regime,
and the position of the bifurcation line is simply given by~:

\begin{equation}
\label{eq:C12}
\left(\frac{K}{W}\right)_{c} = \mu_{\phi} - \mu_{\dot{x}}
\end{equation}
For the RR model, $\mu_{\phi} - \mu_{\dot{x}} = B - A$ is a constant,
$(K/W)_{c}$ is $V$-independent.

At the bifurcation, the two (complex conjugate) neutral modes oscillate
at the critical frequency~:

\begin{equation}
\label{eq:C13}
\Omega_{c} = \frac{V}{D_{0}} \sqrt{\frac{\mu_{\phi}}{\mu_{\dot{x}}} -
1}
\end{equation}
That is, the bifurcation is of the Hopf type. A third order
perturbation expansion can be performed standardly provided
that $\mu_{\phi}$ and $\mu_{\dot{x}}$ are not mere constants (see
$\S$II). It shows that the stick-slip bifurcation is direct - i.e.
that the SS amaplitude grows continuously when decreasing e.g. $K$
below its critical value.

Note finally that, for a velocity-strengthening $\mu_{d}$,
($\mu_{\phi} - \mu_{\dot{x}}) < 0$, the system is always stable
against {\it infinitesimal} perturbations. This does not preclude the
possibility of a {\it finite} amplitude instability.
Brockley \cite{Brockley} has shown that -- in the case where
$\mu_{\phi} = 0$ -- when one takes inertia into account, the system
does exhibit a strongly hysteretic Hopf bifurcation. As explained
in $\S$II, one expects this situation to prevail, for rough-on-rough
MCI, at large enough driving velocities ($V > V_{min}$)
for which geometric aging becomes inactive. A Brockley-like regime
has indeed been observed \cite{TBTrieste}, for a paper/paper interface, in
the cm/sec range.

\section* {Appendix D~: The Schallamach model of adsorption-controlled
friction}

Following Schallamach's seminal article \cite{Schall}, consider an
extended interface between a soft slider and a hard flat track
covered with adhesive sites which can pin the slider molecules.
Represent the slider as a set of $N_{0}$ identical and {\it independent}
chains of stiffness $\kappa$, potentially forming bonds by adsorption
of their end monomer onto the track. Adsorption and subsequent desorption
are thermally activated, and, when sliding, desorption is aided by
advection. The number $N$ of bonds is therefore expected to be a
function of the sliding velocity $V$ and of the temperature $T$. In
steady motion, the elastic force exerted on a given bond increases
linearly with time until the bond snaps off and the stored energy is
dissipated.

The frictional force thus reads~:

\begin{equation}
\label{eq:D1}
F = N \kappa V \bar{t}
\end{equation}
where $\bar{t}$ is the average lifetime of a bond.

Let $\tau$ be the average time for which a chain remains depinned before
readsorbing. The stationary number of bonds is~:

\begin{equation}
\label{eq:D2}
N = N_{0} \frac{\bar{t}}{\bar{t} + \tau}
\end{equation}
and the friction force

\begin{equation}
\label{eq:D3}
F = N_{0}\kappa V \tau \frac{(\bar{t}/\tau)^{2}}{1 + (\bar{t}/\tau)}
\end{equation}

\begin{figure}[h]
     \includegraphics[width = 5cm]{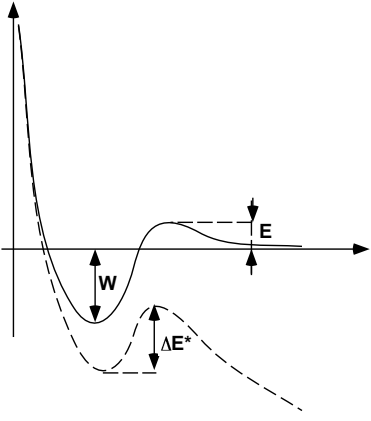}
    \caption{
    \label{fig:pot Schall}
    Sketch of the adhesive pinning potential (full line). When the pinned
    chain is stretched, the desorption barrier decreases from $\Delta E =
    W + E$ to $\Delta E^{*}$ (dashed line).  }
    \end{figure}

In order to desorb (resp. adsorb), a chain must overcome an energy barrier of
height $\Delta E = W + E$ (resp. $E$), as sketched on
Figure \ref{fig:pot Schall}. When the pinned chain is stretched at
velocity $V$, the barrier is lowered and its effective height becomes
$ \Delta E^{*} = \Delta E - \kappa bVt$, where $b$ is a length of
atomic order. Schallamach assumes that the resulting desorption rate
is~:

\begin{equation}
\label{eq:D4}
r(t) = \frac{1}{\tau_{0}} \exp \left[-\frac{\Delta
E^{*}(t)}{k_{B}T}\right]
\end{equation}
where $\tau_{0}^{-1}$ is an attempt frequency, and $t$ the age of the
bond.

Note that this expression for $r$ tacitly assumes that thermal
activation is efficient enough for all bonds to break before reaching
the deterministic threshold such that $\Delta E^{*} = 0$, i.e. that
$\Delta E^{*}(\bar{t}) \gg  k_{B}T$. We will discuss this assumption
in more detail below.

The time $\tau$ is that for thermal activation above the barrier
$E$, of unspecified origin, and Schallamach sets~:

\begin{equation}
\label{eq:D5}
\tau = \tau_{0} \exp \left(E/k_{B}T\right)
\end{equation}
an expression which neglects in particular advection effects.

Consider a set of $n_{0}$ bonds all formed at the same time $t = 0$.
At time $t$, the number of these, $n(t)$, which have not desorbed obeys~:

\begin{equation}
\label{eq:D6}
\frac{dn}{dt} = - r(t) n
\end{equation}

The average bond lifetime $\bar{t}$ then reads, with the help of
equations (\ref{eq:D4}),(\ref{eq:D6})~:

\begin{equation}
\label{eq:D7}
\bar{t} = \int_{0}^{\infty}\, \frac{n(t)}{n_{0}} dt = \tau_{out}
\frac{V_{0}}{V} \int_{0}^{\infty}\,\frac{e^{-y}}{y + (V_{0}/V)} dy
\end{equation}
where
$\tau_{out} = \tau_{0} \exp\left(\Delta E/k_{B}T\right)$ is the
desorption time of the unstretched chain, and
$V_{0} =k_{B}T/\kappa b \tau_{out}$.

This, together with expression
(\ref{eq:D3}), yields the friction force in the steady state. More
precisely, in the small and large velocity limits~:

\begin{equation}
\label{eq:D8}
\bar{t}/\tau_{out} = 1 - \frac{V}{V_{0}} + \ldots\,\,\,\,\,\,\,\,\, (V \ll
V_{0})
\end{equation}

\begin{equation}
\label{eq:D9}
\bar{t}/\tau_{out} = \frac{V_{0}}{V} \left[ \ln \frac{V}{V_{0}} -
0.577 +\ldots\right]\,\,\,\,\,\,\,\,\, (V \gg V_{0})
\end{equation}

\begin{figure}[h]
    \includegraphics[width = 5cm]{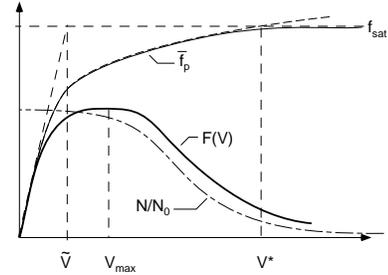}
    \caption{
    \label{fig:tout Schall}
    Qualitative plot of the average pinning force $\bar{f}_{p}$ (thin
    line), the fraction of bonded chains $N/N_{0}$ (dashed line), the
    friction force $F(V)$ (thick line) as predicted by Schallamach's model,
    for $\tau/\tau_{out} \ll 1$.}
    \end{figure}

$\bar{t}$ decreases monotonously with $V$, while the elastic energy
$\kappa bV\bar{t}$ stored before
debonding increases from linearly ($V \ll V_{0}$) to logarithmically
($V \gg V_{0}$) as shown on Figure \ref{fig:tout Schall}. 
Also shown on this figure is the average number of bonds in
steady state $N(V)$, which decreases all the more slowly to zero in
the large $V$ limit that $\tau/\tau_{out}$ is small.

As a result $F(V)$  exhibits a maximum
at $V = V_{max}$ (Figure \ref{fig:tout Schall}) resulting from the
interplay of two effects~:

    -- the decrease with $V$ of the number of bonds.

    -- the increase of the average pinning force $\bar{f}_{p} = \kappa
V\bar{t}$.

In the limiting regimes~:

\begin{equation}
\label{eq:D10}
F(V) \simeq N_{0} \kappa \frac{\tau_{out}^{2}}{\tau + \tau_{out}}
V\,\,\,\,\,\,\,\,\,\,\,\,\,\,\,\,\, (V \ll V_{0})
\end{equation}

\begin{equation}
\label{eq:D11}
F(V) \simeq N_{0} \frac{\tau_{out}}{\tau} \frac{k_{B}T}{b}
\frac{V_{0}}{V} \ln^{2} \frac{V}{V_{0}}
\,\,\,\,\,\,\,\,\,\,\,\,\,\, (V \gg V_{max})
\end{equation}

  Let us now discuss in more detail Schallamach's results and
  assumptions. Two weak points in the theory are immediately clear.

  $\bullet$ It predicts vanishing friction at large $V$, since, in this
  limit, complete depinning is achieved. The junction is then a mere
  sheared liquid layer, whose viscous dissipation has been neglected.
The corresponding contribution $F_{vis}(V)$ to the total friction
force $F_{tot}(V) = F(V) + F_{vis}(V)$
becomes dominant for $V \gg V_{max}$. Its precise expression of
course depends on the nature of the slider material (e.g. hydrogel
{\it vs} elastomer).

  $\bullet$ As mentioned above, Schallamach's expression for the
  desorption rate $r$ is valid only as long as $\Delta E^{*}(\bar{t})
  \ll k_{B}T$. Since equation (\ref{eq:C4}) results in an unboundedly
  growing value of $\kappa bV\bar{t}$ (see Figure \ref{fig:tout
  Schall}), this assumption fails for $V \gtrsim V^{*}$ such that $\Delta
  E = \kappa bV^{*}\bar{t}(V^{*})$, which yields $V^{*} \sim
  V_{0}\exp(\Delta E/k_{B}T)$. In this fast regime, we are back to the
  scenario of $\S$ II.C.4. Advection-controlled deterministic bond
  breaking becomes more efficient than ``premature" thermally activated
  depinning, resulting in a value of $\bar{f}_{p}$ which crosses over
  from the Schallamach logarithmic regime to the deterministic
  saturation value $f_{sat} = \Delta E/\kappa b$. That is, as shown on Figure
  \ref{fig:tout Schall}, $\bar{f}_{p}$ exhibits three regimes~: an
  initial linear one for $V \lesssim \tilde{V}$, a logarithmic
one for
  $\tilde{V} \lesssim V \lesssim V^{*}$, saturation for $V > V^{*}$.
  The lower crossover $\tilde{V}$ can be evaluated as $\tilde{V}
  \simeq V_{0}(\Delta E/k_{B}T) \simeq \Delta E/\kappa b \tau_{out}$.

  Keeping track of these two corrections results in a total friction
  force curve whose qualitative shape depends on the order of
  magnitude of the parameter $\tau/\tau_{out}$. Note that, within
  Schallamach's assumption (equation (\ref{eq:C5})) $\tau/\tau_{out} <
  1$.\\

  If $\tau/\tau_{out} \ll 1$, one easily evaluates $V_{max} \approx
  V_{0}\tau_{out}/\tau$, so that

$$\frac{V}{V_{max}} \approx \frac{\Delta E}{k_{B}T}
  \frac{\tau}{\tau_{out}} \approx\frac{\Delta E}{k_{B}T}
  e^{-W/k_{B}T} \ll 1$$
and $F(V)$ exhibits a wide plateau followed by a slow decrease.
  Whether or not the total force $F_{tot} = F + F_{vis}$ exhibits a
  $\cal N$-shape for $V > V_{max}$ depends on the effective viscosity of
  the unpinned junction and on the density of pinning sites $N_{0}$.

  As $\tau/\tau_{out}$ increases towards $1$, the width of the plateau
  decreases with $V_{max}$, $F(V)$ becomes steeper on the high-$V$
  side, so that the more likely a $\cal N$-shaped $F_{tot}(V)$, hence
  a $V$-weakening friction regime leading to stick-slip.

  Finally, it is worth recalling that, when the system is loaded from
  rest at a prescribed velocity $V_{load}$ larger than, typically,
  $\tilde{V}$, its response
  is initially quasi-elastic, the upper limit of this regime appearing as a
  static threshold. Since $\tilde{V} \sim \Delta E/\kappa b\tau_{out}$,
  reasonable estimates lead to conclude that in order for $\tilde{V}$
  to lie in the sub-$\mu$m/sec range, adsorption
should be quite strong  -- corresponding to binding energies $W$ not far
below the eV level.


\end{document}